%% file: main-white-paper.tex
\documentclass[letterpaper,twocolumn,10pt]{article}
\usepackage{sty/usenix}

\usepackage{sty/macros_general}
\usepackage{sty/macros_zpr}

\begin{document}

\date{}

\title{

{\large \sol: Transmuting Trusted Third-Party Services Into Trustless Atomic Broadcast }

}


\author{
Matteo Bj\"{o}rnsson,
Taylor Hardin,
Taylor Heinecke,
Marcin Furtak,
David L. Millman,
Mike P. Wittie\\
(alphabetical)\\\\
{\rm\missellenfont{BLOCKY}\!\!, Inc.}\\
\ttt{research@blocky.rocks}
}

\maketitle

\input{0.abstract.tex}

\input{1.introduction}

\input{2.services}

\input{3.zpr}

\input{4.implementation}

\input{5.evaluation}

\input{6.discussion}

\input{7.related_work}

\input{8.conclusions}

\input{9.acknowledgments}

\bibliographystyle{plain}
{\small
\bibliography{references/zpr-mwittie,references/publications-mwittie}
}

\ifdraft
\input{10.appendix}
\else
\fi

\end{document}

%% file: 0.abstract.tex
\begin{abstract}
Distributed ledger technologies~(DLTs) rely on distributed consensus mechanisms to reach agreement over the order of transactions and to provide immutability and availability of transaction data.
Distributed consensus suffers from performance limitations of network communication between participating nodes.
BLOCKY \sol\ guarantees immutability, agreement, and availability of transaction data, but without relying on distributed consensus.
Instead, its construction process transfers trust from widely-used, third-party services onto \sol's correctness guarantees.
\sol\ blocks are built by a pipeline of specialized services deployed on a small number of nodes connected by a fast data center network.
As a result, \sol\ transaction throughput approaches network line speeds and block finality is on the order of 500\,ms.
Finally, \sol\ infrastructure creates blocks centrally and so does not need a native token to incentivize a community of verifiers.
\end{abstract}

%% file: 1.introduction.tex
\section{Introduction}


Distributed ledger technologies~(DLTs), or blockchains, make possible a growing number of decentralized alternatives to traditionally centralized services.
Proof-of-Work blockchains made the development of decentralized services only marginally practical due to high delay, low transaction throughput, and high and unpredictable cost of recording and executing transactions.
Newer blockchain proposals address these limitations with novel consensus mechanisms, faster block distribution techniques, and less speculative transaction pricing~\cite{Yakovenko19Solana, Hentschel19Flow, Buterin19EIP1559}.
Fundamentally, however, the performance of blockchains is limited by their reliance on distributed consensus algorithms, which suffer from performance limitations of network communications between participating nodes~\cite{Klarman18bloXroute}.
As a result, some DLTs centralize their infrastructure to improve network performance and provide higher transaction throughput, although at the cost of weaker guarantees.


Blockchains guarantee immutability, agreement, and availability. 
They provide immutability by cryptographically linking blocks in a way that makes their retroactive modification without renewed distributed agreement near impossible.
Agreement, in turn, comes from a blockchain's distributed consensus protocol, which ensures that the creation of new blocks follows preset rules.
Finally, availability comes from the replication of blockchain state among distributed nodes that prevents its deletion and, in the case of public blockchains, provides censorship resistance.
While these mechanisms may seem inextricably linked in blockchains, we demonstrate they they can be uncoupled, so that blockchain benefits do not need bear the cost of consistency in distributed systems~\cite{Brewer12CAP}.


We propose to decouple blockchain mechanisms by revisiting the underlying trust relationships between a blockchain and its users.
Blockchains are considered trustless peer-to-peer systems, because to use them users do not need to trust each other.
We point out, however, that other widely used systems are based on strong, but limited trust relationships.
For example, users generally trust certificate authorities' assertions of public keys.
Similarly, authentication services based on OAuth~2.0~\cite{Hardt12OAuth} are generally trusted to issue correct authentication tokens.
Such trust relationships emerge from a clear self-interest in the correctness of the service by its provider that goes beyond the benefit of circumventing its one particular use case.


In this paper, we demonstrate that strong, but limited trust in third-party services can give rise to alternative mechanisms for immutability, agreement, and availability that do not rely on distributed consensus.
We combine these mechanisms in a blockchain we dub \sol\ that transfers users' trust in these third-party services onto its correctness guarantees.
We consider \sol\ trustless in that it does not introduce trust dependencies beyond the assumption of user trust in the third-party services, on which it is based.
\sol\ runs on a pipeline of specialized services connected by a fast data center network, which allows it to
achieve projected throughput approaching line speeds, 
reach block finality of around 100\,ms,
and operate without a native token to incentivize a community of verifiers.




%% file: 2.services.tex
\section{Trusted Services}
\label{sec:services}

\sol\ makes a departure from distributed blockchain implementations to provide its guarantees of immutability, agreement, and availability based on trusted third-party services.
To ground the presentation of \sol\ in \refsec{sec:zpr}, in this section, we describe the abstractions and implementations of three trusted services, on which \sol\ relies. 
To facilitate further discussion, we summarize the notation introduced throughout this paper in \reftbl{tbl:notation}.

\begin{table}[h]
    \centering
    \small
    \caption{Paper notation.}
    \input{images/notation} 
    \label{tbl:notation}
    \vspace{-10pt}
\end{table}

\subsection{Trusted \timesvclong}
\label{sec:services:timestamp}

The function of the trusted \timesvclong\ is to provide accurate and trustworthy physical clock timestamps to unique objects.
Objects are typically identified by hash values that summarize object bytes.
However, hash collisions remain a possibility, which can result in rare cases of mistaken identity.
To avoid these entirely, we define the \linkf\ function 
\mbox{$\link(\xvar) = \langle \xvar.\uuid, \hash(\xvar) \rangle$},
where
$\xvar$ is an object data structure,
$\xvar.\uuid$\footnotemark is its universally unique identifier~(UUID)~\cite{Leach05UUID},
and
$\hash(\xvar)$ is the hash digest of $\xvar$ using a cryptographic hash function $\hash$ such as \ttt{SHA3_256}.
\footnotetext{We use the ``$.$'' operator for member selection.}
The link function creates a unique reference to $\xvar$; 
while $\xvar.\uuid \in \link(\xvar)$ uniquely identifies $\xvar$, 
$\hash(\xvar) \in \link(\xvar)$ ensures reference integrity.



We assume the existence of a \timesvclong\ $\timesvc$, which maintains a public/private key pair $\key_\timesvc^+/\key_\timesvc^-$ and an accurate physical clock.
A \timesvclong\ provides the following abstract interface:
\begin{equation}
    \notag
    \begin{split}
        \time &\coloneqq \stamp(\link(\xvar)) \\
        \true/\false &\coloneqq \validate(\key, \time) 
    \end{split}
\end{equation}
The \stamp\ function takes an object link tuple $\link(\xvar)$ as input, 
reads the physical clock value~$\ts$, 
computes a UUID~$\uuid$,
and uses these to create a timestamp attestation~$\time$.
The attestation is a tuple 
$\time = \langle \bytes, \ts, \uuid, \sig \rangle$, 
where
$\bytes$ are the bytes of $\link(\xvar)$
and
\mbox{$\sig = \encrypt{\key_\timesvc^-}{\hash(\bytes, \ts, \uuid)}$}
is a cryptographic signature by the private key of $\timesvc$.\footnotemark
\footnotetext{We use the Kerberos security protocol notation~\cite{Steiner98Kerberos, Briais05Formal}.} 
The \validate\ function takes a public key $\key$ and a timestamp attestation $\time$ as input to determine that $\bytes$, $\ts$, and $\uuid$ are correctly signed, or that
$\hash(\time.\bytes, \time.\ts, \time.\uuid) = \decrypt{\key}{\time.\sig}$, 
where $D$ is a decryption function.
When users trust a \timesvclong\ and $\validate(\key_\timesvc^+, \time)$ returns \true, they can trust that the \timesvclong\ $\timesvc$ has witnessed bytes $\time.\bytes$ at time $\time.\ts$.
Note that while the \stamp\ function must execute on the trusted \timesvclong, the \validate\ function may be run by a user as long as $\key_\timesvc^+$ is well-known.

We implement the \timesvclong\ based on a user authentication service, such as AWS Cognito or Auth0 Login~\cite{2022AmazonCognito, Auth024Embedded}.
A user authentication service accepts user credentials (\ttt{username} and \ttt{password}), 
validates the credentials, and produces an OAuth Identity token\cite{openID}.
OAuth Identity tokens follow the JSON Web Token~(JWT) standard and JWTs provide claims that can only be verified using the public key of the service that created the token.
An OAuth Identity token must provide the ``issued at time'' \ttt{iat} claim, or the timestamp of token creation.
The user authentication service also adds the \ttt{email} claim, to represent the user's \ttt{username}, and the ``JWT ID'' \ttt{jti} claim, or the token's UUID.

We can use a user authentication service as a \timesvclong\ as follows. 
The \stamp\ function converts the input link tuple $\link(\xvar)$ to an email address
\mbox{$\texttt{email} = \hex(\hash(\xvar)) \,\texttt{@}\,\, \hex(\xvar.\uuid) \texttt{.bky.sh}$},
where \hex\ creates a hexadecimal string representation of a byte array.
This approach bakes in the component values of $\link(\xvar)$ into an email address, while remaining compliant with email address length limits defined in RFC~5321~\cite{Klensin08SMTP}.
According to RFC~5321 the ``local part'' of the address (preceding \ttt{@} sign) is restricted 64 bytes, which is the exact length of a 256-bit, or a 32-byte hash value represented as a hexadecimal string.
Further, the ``path'' of the address (following the \ttt{@} sign) is restricted to 256 bytes, which is more than enough for a hexadecimal representation of a UUID followed by ``\texttt{.bky.sh}''. 
Finally, we authenticate the user, which produces an OAuth Identity Token.
Given the well-known public key of the user authentication service, a client can validate this token and parse out its claims.
Therefore, if a user trusts the user authentication service, the user trusts that the bytes 
$\hash(\xvar)$ and $\xvar.\uuid$, represented in the token's \ttt{email} claim, existed at time $\time.\ts$ specified in the \ttt{iat} claim.

\subsection{Trusted \seqsvclong}
\label{sec:services:sequence}

The function of the trusted \seqsvclong\ is to provide consecutive sequence numbers to distinct events.
A \seqsvclong\ $\seqsvc$ maintains a public/private key pair $\key_\seqsvc^+/\key_\seqsvc^-$ and 
a counter that starts at $0$ increments by $1$. 
A \seqsvclong\ provides the following abstract interface:
\begin{equation}
    \notag
    \begin{split}
        \seq &\coloneqq \sequence(\bytes) \\
        \true/\false &\coloneqq \check(\key, \seq)  
    \end{split}
\end{equation}
The \sequence\ function takes an object link tuple $\link(\xvar)$ as input, where $\xvar$ is an object representation of a unique event. 
On invocation, the \sequence\ function increments the counter by one, records the counter value as $\ctr$, and produces a sequence attestation 
\mbox{$\seq = \langle \bytes, \ctr, \sig \rangle$,} 
where
$\bytes$ are the bytes of $\link(\xvar)$
and
$\sig = \encrypt{\key_\seqsvc^-}{\hash(\bytes, \ctr)}$ is a cryptographic signature.
The \check\ function takes a key $\key$ and a sequence attestation $\seq$ as input to determine that $\bytes$ and $\ctr$ are correctly signed, or that
$\hash(\seq.\bytes, \seq.\ctr) = \decrypt{\key}{\seq.\sig}$.
When users trust a \seqsvclong\ and $\check(\key_\seqsvc^+, \sig)$ returns \true, they can trust that a unique event represented by $\seq.\bytes$ was witnessed by the \seqsvclong\ $\seqsvc$ as the $\seq.\ctr^\text{th}$ event.
Similarly to \stamp, the \check\ function may be executed by a user with a well-known $\key_\seqsvc^+$.

\begin{figure}[h]
    \centering
    \includegraphics[scale=0.75]{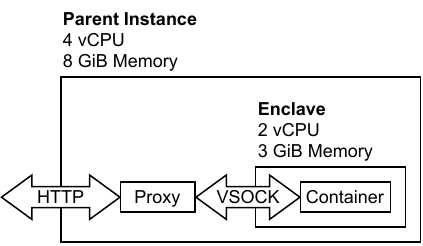}
    \caption{AWS Nitro Enclave.}
    \label{fig:ne}
\end{figure}

We implement a \seqsvclong\ based on the security and correctness guarantees provided by the AWS Nitro Enclaves trusted execution environment~(TEE)~\cite{NitroEnclaves}.
Nitro Enclaves create an isolated execution environment inside an Elastic Cloud Compute~(EC2) instance based on the same Nitro Hypervisor technology that provides isolation between EC2 instances themselves~\cite{2022NitroSecurity}.
Inside an EC2 parent enclave, as shown in \reffig{fig:ne}, an enclave runs a container on its own kernel, memory, and vCPU resources sequestered from the parent instance.
An enclave has no persistent storage, interactive access, or external networking.
The only means for the parent instance to interact with an enclave is through a VSOCK socket.
The parent instance, however, may proxy external requests for example by running a Hypertext Transfer Protocol~(HTTP) server.
Finally, applications running inside the enclave may request attestations from the Nitro Secure Module~(NSM)~\cite{22NSM}.
An attestation $\encl$ includes information about the enclave environment recorded as
hashes 
of the parent instance ID, the enclave image file~(container) as $\encl.\image^\hash$, and the application requesting the attestation.
Optionally, the attestation may also include the public key of the application as $\encl.\key$ and up to 1024\,B of user data~\cite{22NitroUserGuide}.
NSM packages the attestation as a Concise Binary Object Representation~(CBOR)-encoded, CBOR Object Signing and Encryption~(COSE)-signed object by the AWS Nitro Attestation key pair $\key^+_\enclsvc/\key^-_\enclsvc$, where $\key^+_\enclsvc$ is in a well-known root certificate~\cite{22NitroRootCert}.

We implement a trusted \seqsvclong\ on a Nitro Enclave as follows.
The parent instance runs an API gateway, which exposes the \seqsvclong\ \sequence\ function to clients and proxies calls to it to the enclave~$\seqsvc$.
A \sequence\ request includes bytes $\bytes$ representing a unique event ID, which serves to identify distinct idempotent requests to~$\seqsvc$.
Upon receiving a \sequence\ request the enclave increments an in-memory counter and produces a sequence attestation
\mbox{$\seq = \langle \bytes, \ctr, \sig \rangle$} as described earlier.
It is important to note that repeated requests to sequence the same $\bytes$ will not increment the counter and produce a new sequence attestation; instead, the \seqsvclong\ will serve a previously computed $\langle \bytes, \ctr, \sig \rangle$.
Similarly to a timestamp attestation, we represent $\seq$ as a signed JWT, with custom claims representing $\bytes$ and $\ctr$ and the ``key id'' \ttt{kid} header field storing $\hash(\key^+_\seqsvc)$.
For convenience we use $\seq.\key$ to refer to \ttt{kid} in the JWT representing $\seq$.

To demonstrate to clients that the sequence attestation $\seq$ comes from a \seqsvclong\ $\seqsvc$ running inside a Nitro Enclave, the \seqsvclong\ requests an enclave attestation over its public key $\key^+_\seqsvc$.
The NSM produces an enclave attestation $\encl$, whose public key field \mbox{$\encl.\key = \key^+_\seqsvc$}.
We make the enclave attestation $\encl$ well-known, so that users can use $\encl.\key$ to check multiple sequence attestations from $\seqsvc$.

Upon receiving the sequence attestation $\seq$, a client can check it locally.
First, the client verifies the authenticity of the well-known enclave attestation $\encl$ against the well-known key $\key^+_\enclsvc$.
Second, the client calls $\check(\encl.\key, \seq)$ locally and when it returns \true\ the client can trust that event $\seq.\bytes$ was assigned the sequence number $\seq.\ctr$ by a \seqsvclong\ $\seqsvc$.
It is important to note that in our \seqsvclong\ implementation the enclave generates the key pair $\key_\seqsvc^+/\key_\seqsvc^-$ on startup, which is distinct across all enclave instantiations.
Consequently, every sequence attestation produced by an enclave is unique since the enclave uses a distinct $\key_\seqsvc^-$ to sign an incremented $\ctr$.
As a result, it is not possible, even if a \seqsvclong\ is restarted, to produce two sequence attestations with the same $\ctr$ for different $\bytes$ signed by $\key_\seqsvc^-$.

Finally, the last issue is that of user trust.
A user may trust that the AWS~Nitro~Enclaves system works correctly.
The \seqsvclong, however, is based on our implementation.
We will make it possible for anyone to inspect our \seqsvclong\ implementation at \url{https://github.com/blocky/sequencer}.
The implementation's code and build tools are publicly available and so a user may perform an audit of the code, build an image $\image$, and take the hash of the image $\hash(\image)$.
Recall that the enclave attestation $\encl$ contains a sighed hash of the image running on the enclave as $\encl.\image^\hash$. 
Therefore when a user trusts an implementation of \seqsvclong, the user can verify that a sequence attestation was generated using the trusted implementation by checking that $\hash(\image) = \encl.\image^\hash$.
That is, the hash of the image that the user built matches the hash of the image in the enclave attestation.

\subsection{Trusted \repsvclong}

The function of the trusted \repsvclong\ is to provide stable storage to data objects.
Since storage nodes have non-zero mean time between failures~(MTBF), storage stability is probabilistic and comes from replication of data objects among nodes with mostly independent failures.
A \repsvclong\ provides the following abstract interface:
\begin{equation}
    \notag
    \begin{split}
        &\replicate(\xvar) \\
        \xvar \coloneqq\,\, &\fetch(\link(\xvar))  
    \end{split}
\end{equation}
The \replicate\ function takes as input object $\xvar$ replicates its bytes across storage nodes under the bytes of $\link(\xvar)$ as the retrieval key.
The \fetch\ function takes the link tuple $\link(\xvar)$ as input and returns the object $\xvar$ loaded from one or more replicas.

Fundamental models of distributed systems commonly assume that nodes have access to stable storage to imply that protocol data survives node failures.
To argue the correctness of \sol, we need to both strengthen the notion of stability and make it more specific.
We define a \emph{stable storage service} as providing durability, immutability, and verifiability.
Durability means that a stored object will remain eventually accessible.
Immutability means that a stored object will not change in storage.
Verifiability means that a third party may verify that a storage service provides durability and immutability.

We implement the \repsvclong\ based on the correctness guarantees of Write Once, Read Many~(WORM) cloud storage systems.
WORM systems enable their clients to protect data stored in buckets against inadvertent modification, or deletion, to meet regulatory requirements on data retention~\cite{03SEC}.
Specifically, the AWS Simple Storage Service~(S3), Microsoft Azure Blob Storage, and Google Cloud Storage guarantee object immutability through legal/compliance holds on data objects that prevent anyone, including the bucket owner, from deleting or modifying objects~\cite{Amazon2022S3UserGuide, Microsoft21Legal, Google22Retention}.
To protect against accidental data loss, Amazon, Microsoft, and Google design these services to provide 11-nines of durability by replicating data across availability zones within a region~\cite{Amazon2022S3UserGuide, Microsoft22AzureDurability, Google21Durability}.
Formally, 11-nines of durability guarantee that a stored object will remain accessible over a year with the probability greater than $1 - 10^{-11}$.

Finally, cloud storage providers allow, with some custom configuration, to make bucket settings publicly readable, which allows anyone to verify that object holds are enabled.
Alternatively, an AWS Nitro Enclave with read-only credentials, may inspect bucket settings and emit publicly verifiable attestations over the bucket's setting to demonstrate that objects are in compliance mode.

Although the 11-nines durability guarantee is the industry standard, it is not sufficient by itself for storage of \sol\ objects.
As we detail in \refsec{sec:implementation}, \sol\ stores four objects per block.
We intend to push \sol\ performance to 10 blocks per second.
If we assume that we want a \sol\ blockchain to operate for 100 years, it would generate $1.26\!\x\!10^{11}$ objects.
For a \sol\ blockchain to remain viable all these objects must remain accessible.
Backblaze presents details of how cloud storage providers calculate durability~\cite{Wilson18Backblaze}.
We extend their analysis to calculate the joint durability of a set of objects, or the probability that all the objects retain durability over a period of time.
We calculate the joint durability of \sol\ objects stored on a single cloud bucket over 100 years to $4.06\!\x\!10^{-28}$ -- a woefully insufficient number.

To increase the probability that a \sol\ chain remains viable, we apply erasure coding to \sol\ objects to distribute their storage.
Using random linear network coding~(RLNC)~\cite{Ho2006Random}, we encode each object into $6$ shards and need any $3$ to decode the object.
We partition these shares across buckets located in six regions, two each for AWS~S3, Azure Blob Storage, and Google Cloud Storage, which allows us to assume independence of bucket failures.
We use the binomial cumulative distribution function to calculate the durability of a \sol\ object to 35-nines and the cumulative durability of a \sol\ chain over a 100 years to 14-nines~\cite{BLOCKY23ZipperChainReplication}.

\subsection{Reliability of Trusted Services}
\label{sec:services:reliability}

In addition to being trusted, the design of \sol\ also requires the \timesvclong, \seqsvclong, and \repsvclong\ to be reliable in that they can recover from crash failures.
While reliability at the cost of temporary unavailability may be assumed for a user authentication service and cloud storage services, the same cannot be done for an AWS~Nitro~Enclave.
An enclave may crash, but because it relies on an internally generated key pair and in-memory state to provide unique sequence attestations, the enclave may not be restarted.
Non-trivial software is inevitably subject to bugs, so it may also be necessary to switch a \sol\ onto another \seqsvclong\ node running an updated version of its software.
To address these issues, we have designed a mechanism for \sol\ to reliably replace  \seqsvclong\ nodes.
We make the description of this mechanism a focus of a separate paper. 
For the remainder of this paper, however, we assume that the \seqsvclong\ is reliable, at the cost of no new \sol\ blocks being possible in the case of a \seqsvclong\ failure.

%% file: images/notation.tex
\begin{tabular}{l|l|l}
    \hline
    Symbol & Meaning & Format/Type\\
    \hline
    $\encl$ & Enclave attestation                   & $\langle \key^+_\seqsvc, \sig, \dots \rangle$\\
    $\enclsvc$ & Enclave Service                    & AWS Nitro Enclave\\

    $\block$ & Block                                & $\langle \uuid, \merkle^\link\!, \,\time^\link \rangle$\\ 
    $\genesis{\block}$ & Genesis block              & Well-known \\
    $\blocks$ & Set of blocks                      & $\{\block_0, \block_1, \dots\}$\\
    
    $\ctr$ & Counter value                          & \tttt{uint}\\ 

    $\tx$ & Transaction representation                        & $\langle \texttt{schema}, \texttt{type}, \link(\texttt{tx}) \rangle$\\
    
    $\tri$ & Triad                                  & $\langle \block, \time, \seq \rangle$ \\

    $\cert$ & Certificate                           & $\langle \tx^\hash\!\!, \genesis{\block}^\link\!\!, \ts, \heightval, \rank  \rangle$ \\

    $\sig$ & Signature of object $\xvar$            & $\encrypt{\key^-}{\hash(\xvar)}$\\

    $\heightval$ & Height of triad -- $\height(\tri)$         & See \refsec{sec:zpr:correctness}\\

    $\hash(\xvar)$ & Hash function                  & $\xvar^\hash = $ \tttt{SHA3\_256(}$\xvar$\tttt{)}\\



    $\key^+/\key^-$ & Public/private key            & \tttt{byte[]} / \tttt{byte[]}\\

    $\link(\xvar)$ & Link function                  & $\xvar^\link = \langle \xvar.\uuid, \hash(\xvar) \rangle$ \\  

    $\merkle$ & Merkle tree                         & $\langle \uuid, \leaves \rangle$ \\
    $\merkles$ & Set of Merkle trees                & $\{\merkle_0, \merkle_1, \dots\}$\\

    $\ts$ & Physical clock reading                  & \tttt{uint}\\

    $\seq$ & Sequence attestation                   & $\langle \bytes, \ctr, \sig \rangle$\\
    $\seqs$ & Set of sequence attestations          & $\{\seq_0, \seq_1, \dots\}$\\
    
    $\seqsvc$ & \seqsvclong                         & Sequencer on $\enclsvc$\\

    $\time$ & Timestamp attestation                 & $\langle \bytes, \ts, \uuid, \sig \rangle$\\
    $\times$ & Set of timestamp attestations        & $\{\time_0, \time_1, \dots\}$\\
    $\timesvc$ & \timesvclong                       & AWS Cognito, etc.\\ 
    $\rank$ & Rank of txn. in a Merkle tree    & See \refsec{sec:zpr:correctness}\\

    $\uuid$ & Universally unique identifier         & \tttt{byte[16]}\\
    $\leaves$ & Ordered set of Merkle leaves        & \tttt{byte[][]}\\
    $\bytes$ & Byte array                           & \tttt{byte[]}\\

    $\encrypt{\key}{\bytes}$ & Encryption of $\bytes$ with $\key$   & \tttt{byte[]}\\ 
    $\decrypt{\key}{\bytes}$ & Decryption of $\bytes$ with $\key$   & \tttt{byte[]}\\ 

    \hline
\end{tabular}

%% file: 3.zpr.tex
\section{\sol}
\label{sec:zpr}

We propose a new blockchain we dub \sol\ that provides immutability, agreement, and availability based on strong, but limited user trust in the Timestamp, Sequence, and Replication services.
The key innovation of \sol\ is its structure and construction process that transfer user trust in these services onto \sol's correctness guarantees.

\begin{figure}[h]
    \centering
    \input{images/zipperchain}
    \caption{A \sol\ blockchain formed through links between blocks ($\block$), Merkle trees ($\merkle$), timestamps attestations~($\time$), and sequence attestations ($\seq$).}
    \label{fig:zipperchain}
\end{figure}
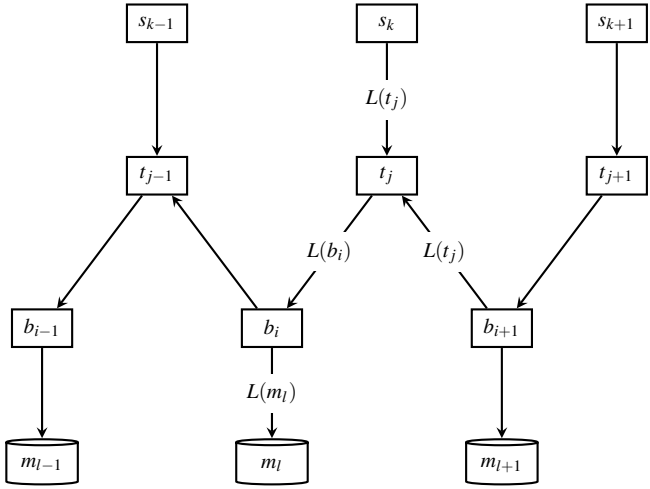

\subsection{Structure}

At its core, a \sol\ blockchain serves to order an incrementally growing set of transactions.
We assume that each transaction $\xvar$, in addition to other fields, contains a unique id $\xvar.\uuid$, so that it is possible to compute the link $\link(\xvar)$.
In the context of \sol, we represent transaction data link $\tx$ as 
$\tx = \langle \texttt{schema}, \texttt{type}, \link(\xvar) \rangle$, 
where \texttt{schema} is a prefix-free key to the definition of the structure of the tuple $\tx$,
\ttt{type} is the type of the transaction,
and $\link(\xvar)$ is the link to identify transaction data stored offchain.
The \ttt{type} field allows a user to selectively process, in order, only transactions of that type, without the need to fetch all transactions from offchain storage.

\reffig{fig:zipperchain} shows the structure of a \sol\ blockchain.
We define a Merkle tree $\merkle$ as $\langle \uuid, \leaves \rangle$, where
$\uuid$ is a distinct UUID assigned during Merkle tree creation
and
$\leaves$ is an ordered set of leaves. 
Each leaf in $\merkle.\leaves$ is a transaction data link $\tx$ 
and the root of a Merkle tree $\hash(\merkle.\leaves)$ summarizes these transactions.\footnotemark
\footnotetext{We use the function ``$\hash$'' to refer to both the process of computing a single hash, such as a \texttt{\footnotesize SHA3-256}, or the process of computing a Merkle tree root, depending on the context.}

We define a block $\block$ as 
$\langle \uuid, \merkle^\link, \time^\link \rangle$, where
$\uuid$ is a distinct UUID assigned during block creation,
$\merkle^\link = \link(\merkle)$ is the link to a Merkle tree, and 
$\time^\link = \link(\time)$ is link to a~(preceding) timestamp attestation.\footnotemark
\footnotetext{
\edit{
The structure 
$\block = \langle \uuid, \merkle^\link, \time^\link \rangle$
represents the common case of a ``data'' block.
In practice, we also use ``control`` blocks, where the link $\merkle^\link$ points not to a Merkle tree, but to a control structure.
The implementation distinguishes between data and control blocks by deserializing the addressee of  $\merkle^\link$ and inspecting its type.
A control structure specifies a \seqsvclong\ and \timesvclong\ by their public keys $\key_\seqsvc^+$ and $\key_\timesvc^+$ and, potentially, other protocol parameters that may be added over time.
For simplicity, we say that a control block $\block$ contains $\block.\key_\seqsvc^+$ and $\block.\key_\timesvc^+$.
}
}

Recall the definition of a timestamp attestation $\time$ in \refsec{sec:services:timestamp} as $\langle \bytes, \ts, \uuid, \sig \rangle$, where
$\bytes$ is a set of bytes representing an object link,
$\ts$ is a physical clock timestamp,
$\uuid$ is a UUID,
and $\sig$ is the signature of the \timesvclong.
We create a timestamp attestation $\time$ over a block $\block$ by calling 
$\time := \stamp(\link(\block))$.

Finally, recall the definition of a sequence attestation $\seq$ in \refsec{sec:services:sequence} as $\langle \bytes, \ctr, \sig \rangle$, where 
$\bytes$ is a set of bytes representing an object link,
$\ctr$ is the counter value, and
$\sig$ is the signature of the \seqsvclong.
We create a sequence attestation $\seq$ over a timestamp attestation $\time$ by calling \mbox{$\seq := \sequence(\link(\time))$}.

To illustrate the relationship between Merkle trees, blocks, timestamp attestation, and sequence attestations we refer again to \reffig{fig:zipperchain}.
Block $\block_\idx$ references the specific set of transactions contained in Merkle tree $\merkle_\idxl$ by setting \mbox{$\block_\idx.\merkle^\link = \link(\merkle_\idxl)$}.
Since $\merkle_\idxl.\uuid \in \block_\idx.\merkle^\link$, the block $\block_\idx$ uniquely identifies the Merkle tree $\merkle_\idxl$ and, 
since $\hash(\merkle_\idxl) \in \block_\idx.\merkle^\link$, the block $\block_\idx$ ensures that the Merkle tree $\merkle_\idxl$ retains integrity.
The timestamp attestation $\time_\idxj$ attests the block $\block_\idx$ by setting
$\time_\idxj.\bytes = \link(\block_\idx)$.
Since $\block_\idx.\uuid \in \time_\idxj.\bytes$, the time attestation $\time_\idxj$ uniquely identifies the block $\block_\idx$ and, 
since $\hash(\block_\idx) \in \time_\idxj.\bytes$, the time attestation $\time_\idxj$ ensures that the block $\block_\idx$ retains integrity.
The sequence attestation $\seq_\idxk$ attests the timestamp attestation $\time_\idxj$ and by setting \mbox{$\seq_\idxk.\bytes = \link(\time_\idxj)$}.
As a result, $\seq_\idxk$ uniquely identifies $\time_\idxj$ since $\time_\idxj.\uuid \in \seq_\idxk.\bytes$ and ensures the integrity of $\time_\idxj$ since $\hash(\time_\idxj) \in \seq_\idxk.\bytes$.
Moreover $\seq_\idxk$ attests $\block_\idx$ transitively through~$\time_\idxj$ to uniquely associate the sequencer number $\seq_\idxk.\ctr$ to the block $\block_\idx$ and extend the surjection from sequencer numbers to blocks.
Lastly, block $\block_{\idx+1}$ uniquely identifies $\time_\idxj$ since $\time_\idxj.\uuid \in \block_{\idx+1}.\time^\link$ and ensures the integrity of $\time_\idxj$ since $\hash(\time_\idxj) \in \block_{\idx+1}.\time^\link$.

Given a block $\block$, timestamp attestation $\time$, and sequence attestation $\seq$, we call 3-tuple $\langle \block, \time, \seq \rangle$ a \emph{triad}.
Given a triad $\tri$, timestamp service public key $\key_\timesvc^+$, and sequencer service public key $\key_\timesvc^+$, we call $\tri$ a \mbox{\emph{true triad for keys $\key_\timesvc^+$ and $\key_\seqsvc^+$}} 
when the following \truetri\ function, defined in \refalg{alg:truetri}, returns $true$.

\begin{algorithm}
\caption{$\truetri\left(\tri, \key_\timesvc^+, \key_\seqsvc^+\right)$}
\label{alg:truetri}
\begin{algorithmic}[1]
\State $cond_1 \gets \tri.\time.\bytes = \link(\tri.\block)$
\label{alg:truetri:bhlink}
\Comment{$\tri.\time$ attests $\tri.\block$}

\State $cond_2 \gets \validate(\key_\timesvc^+, \tri.\time)$
\label{alg:truetri:valid}
\Comment{$\tri.\time$ is valid}

\State $cond_3 \gets \tri.\seq.\bytes = \link(\tri.\time)$
\label{alg:truetri:seqlink}
\Comment{$\tri.\seq$ attests $\tri.\time$}

\State $cond_4 \gets \check(\key_\seqsvc^+, \tri.\seq)$
\label{alg:truetri:check}
\Comment{$\tri.\seq$ is valid}

\State \Return $\bigwedge\limits_{\xvar=1}^4 cond_\xvar$
\Comment{Are all conditions true?}
\end{algorithmic}
\end{algorithm}

When clear from context, we use the term \emph{true triad} and omit referencing the keys for which the triad is true.
For example, assuming that $\validate$ and $\check$ are true for all timestamp and sequence attestations in \reffig{fig:zipperchain}, the triad $\langle \block_\idx, \time_\idxj, \seq_\idxk \rangle$ is a true triad.
On the other hand, the triad $\langle \block_\idx, \time_{\idxj+1}, \seq_k \rangle$ is not a true triad because it fails the condition on line~\ref{alg:truetri:bhlink} in~\refalg{alg:truetri}.
Since a true triad's sequence attestation uniquely identifies its timestamp attestation and, transitively, its block, we address a true triad by the index of its sequence attestation, i.e.,
$\langle \block_\idx, \time_\idxj, \seq_\idxk \rangle = \tri_\idxk$.

\subsection{Initialization}

The identity of a blockchain, starting with Bitcoin, is tied to a well-known hash of the genesis block~\cite{Nakamoto09BitcoinCode}.
We identify a \sol\ blockchain by $\link(\genesis{\block})$, where $\genesis{\block}$ is an genesis block.\footnotemark
\footnotetext{
A $\genesis{\block}$ does not have a previous timestamp attestation, and so $\genesis{\block}.\time^\link = \varnothing$.}
Unlike in Bitcoin, the value of $\link(\genesis{\block})$ is not hard-coded, but passed in by a user into the \sol\ verification process described in \refsec{sec:implementation:verify}.
\edit{
A \sol\ $\genesis{\block}$ is a control block and specifies the chain's \seqsvclong\ and \timesvclong\ as $\genesis{\block}.\key_\seqsvc^+$ and $\genesis{\block}.\key_\timesvc^+$.
}
This mechanism allows multiple \sol\ chains to coexist and gives users the option to select the chain they consider canonical in their context.

To initialize a \sol\ chain we create a genesis triad~$\genesis{\tri}$ that
is a true triad,
contains a unique genesis block, 
i.e., $\genesis{\tri}.\block = \genesis{\block}$, 
and a sequence attestation with counter value 0,
i.e., $\genesis{\tri}.\seq.\ctr = 0$.
A \sol\ chain $\link(\genesis{\block})$ is ready to record transactions once the elements of its genesis triad have been replicated by the trusted \repsvclong.

\subsection{Correctness Guarantees}
\label{sec:zpr:correctness}

The structure of a \sol\ blockchain guarantees immutability, agreement, and availability.
We formally define these three properties below and provide intuitions for how \sol\ guarantees them.\footnotemark
\footnotetext{We are currently working on formal proofs of immutability, agreement, and availability in \sol\ and will present them in a future version of this paper as an Appendix.}

\begin{definition}[Immutability]
If one valid client sees transaction $\tx_\idx$ as preceding transaction $\tx_\idxj$ in a \sol\ ledger, 
then it will continue to see transaction $\tx_\idx$ as preceding transaction $\tx_\idxj$.
\end{definition}

\sol\ guarantees immutability by the alternating structure of blocks and timestamp attestations that come together like the teeth of a zipper as a \sol\ chain grows.
A timestamp attestation provides a third-party trusted signature over the hash of the block, which includes the Merkle root constructed from a set of transaction data, also referred to as transactions.
Any change to these transactions would be detectable as a mismatch between the block hash and the bytes signed by the timestamp attestation.
Thus, as long a timestamp attestation remains a part of a chain, its block remains immutable.

A timestamp attestation remains in the chain, because the following block includes its hash.
Thus an attestation at the at end of a chain signs its block and, transitively, the previous attestation and, indirectly, its block, and so on.

A user that considers a block as belonging a \sol\ blockchain can be sure of the block's integrity by verifying its timestamp attestation.
The user can also check the integrity of the chain by verifying the preceding blocks and attestations all the way to a well-known block zero for a particular \sol\ instance.

\begin{definition}[Agreement]
If one valid client sees transaction $\tx_\idx$ as preceding transaction $\tx_\idxj$ in a \sol\ ledger, then all valid clients see transaction $\tx_\idx$ as preceding transaction $\tx_\idxj$.
\end{definition}

\begin{figure}
    \centering
    \input{images/fork}
    \caption{A fork in a \sol\ blockchain.}
    \label{fig:fork}
\end{figure}
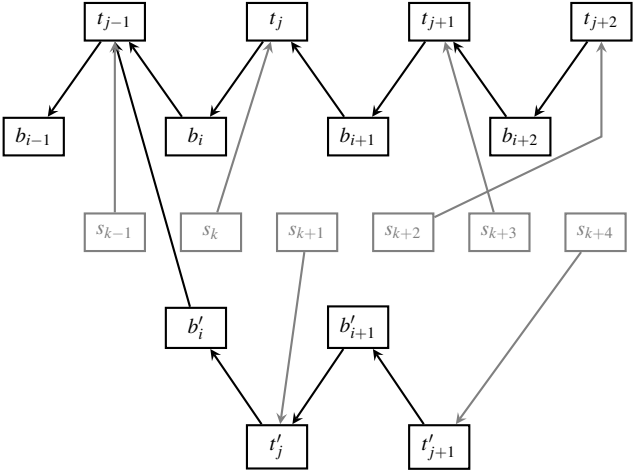

\sol\ guarantees agreement by detecting and eliminating forks, so that all users see the same totally ordered set of blocks and, by extension, transactions.
We define a fork as the existence of two blocks $\block_\idx$ and $\block'_\idx$ with a common immediate predecessor block.
\reffig{fig:fork} illustrates a fork in \sol, where the blocks $\block_\idx$ and $\block'_\idx$ point to the same previous timestamp attestation $\time_{\idxj-1}$ and transitively to a common immediate predecessor block $\block_{\idx-1}$.

Forks, in general, are problematic in blockchains because they create the possibility of an inconsistency of the application state represented on the blockchain.
Let us assume that two users submit transaction data $\tx$ and $\tx'$ that are incompatible with each other.
If there is no fork in the chain, any client reading the blockchain sees the same consistent history in which, say, $\tx$ precedes $\tx'$ in the same block, or across different blocks.
Then according to the rules of an application, the transaction $\tx$ may be applied to the state and $\tx'$ may be ignored.
If, on the other hand, $\tx$ and $\tx'$ are in the forked blocks $\block_\idx$ and $\block'_\idx$ it is not clear how to order $\tx$ and $\tx'$ to determine which transaction to apply and which to ignore.

Forks also create another problem unique to \sol. 
A user verifying the integrity of a \sol\ blockchain by walking through it backwards from a given block, may not know whether the encountered blocks are on the main chain, or on a fork.
This uncertainty is problematic, because transactions in blocks on a fork will not be considered valid by the users following the main chain.
Additionally, the forked blocks may be maliciously deleted (assuming a catastrophic failure of \repsvclong) without users on the main chain detecting that deletion.

\sol\ detects forks by using block heights and sequence attestations to create a total order of blocks.
We define the height of a triad $\height(\tri)$ as the number of triads on the path from $\tri$ to the genesis triad $\genesis{\tri}$.
Given two true triads $\tri$ and $\tri'$ we define an order relation `$\rightarrow$' as 
\begin{equation*}
\begin{split}
    \tri \rightarrow \tri' \iff 
    & \height(\tri) < \height(\tri') ~~ \lor \\
    ( & \height(\tri) = \height(\tri') \land \tri.\seq.\ctr < \tri'\!\!.\seq.\ctr ).
\end{split}
\end{equation*}

In \reffig{fig:fork}, we make the assumption that the sequence attestation suffix represents its counter value, i.e. \mbox{$\seq_\idxk.\ctr = \idxk$}, \mbox{$\seq_{\idxk+1}.\ctr = \idxk+1$}, etc.
With that assumption the total order of true triads in \reffig{fig:fork} is
$$\tri_{\idxk-1} \!\rightarrow\! 
\tri_{\idxk} \!\rightarrow\!
\tri_{\idxk+1} \!\rightarrow\!
\tri_{\idxk+3} \!\rightarrow\!
\tri_{\idxk+4} \!\rightarrow\!
\tri_{\idxk+2}$$ or 
\begin{equation*}
\begin{split}
&\langle \block_{\idx-1}, \time_{\idxj-1}, \seq_{\idxk-1} \rangle \!\rightarrow\!
\langle \block_{\idx}, \time_{\idxj}, \seq_{\idxk} \rangle \!\rightarrow\! 
\langle \block'_{\idx}, \time'_{\idxj}, \seq_{\idxk+1} \rangle \!\rightarrow\! \\ 
&\langle \block_{\idx+1}, \time_{\idxj+1}, \seq_{\idxk+3} \rangle \!\rightarrow\! 
\langle \block'_{\idx+1}, \time'_{\idxj+1}, \seq_{\idxk+4} \rangle \!\rightarrow\! 
\langle \block_{\idx+2}, \time_{\idxj+2}, \seq_{\idxk+2} \rangle.
\end{split}
\end{equation*}

When there is a fork, we determine the true triads on the \emph{main chain} using two rules.  
First, a true triad may be on the main chain only if its predecessor triad is on the main chain.
Second, if multiple true triads have a predecessor on the main chain, the lowest-order true triad is on the main chain.
For example, in \reffig{fig:fork}, assume that $\tri_{\idxk-1}$ is on the main chain and a fork starts with its successors.
Given $\tri_{\idxk}$ and $\tri_{\idxk+1}$ at the start of the fork, we determine which true triad is on the main chain by observing that
\mbox{$\tri_{\idxk}  \rightarrow \tri_{\idxk+1}$}, which implies (deterministically) that $\tri_{\idxk}$ is on the main chain, while $\tri_{\idxk+1}$ is not.
Similarly, since $\tri_{\idxk+3}$ is on the main chain, users can deterministically decide that $\tri_{\idxk+2}$ is on the main chain even though
$\tri_{\idxk+4} \!\rightarrow\! \tri_{\idxk+2},$
because $\tri_{\idxk+2}$'s predecessor is on the main chain, while the predecessor of $\tri_{\idxk+4}$ is not.
Given these rules, the total order of triads on the main chain in \reffig{fig:fork} is
$\tri_{\idxk-1} \!\rightarrow\! 
\tri_{\idxk} \!\rightarrow\!
\tri_{\idxk+3} \!\rightarrow\!
\tri_{\idxk+2}$.
Notice that the sequence attestation counter values in main chain triads do not need to be monotonic.

\begin{definition}[Availability]
If one valid client can obtain a triad $\tri$ then another valid client, with high probability, will be able to obtain the triad $\tri$.
\end{definition}

\sol\ provides strong, probabilistic guarantees on the availability of data by using a trusted Replication Service to increase the distribution of blocks, timestamp attestations, sequencer attestations, and the leaves of the Merkle trees. 
Data replication with a trusted Replication Service creates redundant shards of data distributed among independently failing replicas.
As a consequence, the encoded data remains available to users, even if some of the replicas become unavailable.
Even in the case when a sufficient number of replicas is not momentarily available, users remain confident that the stored data remains intact, because of the high durability guarantees of a trusted Replication Service, and will become available again.
It is important to note that \sol\ records transaction data in a manner that allows users to verify immutability and agreement over blockchain transactions as long as the guarantees of a trusted Replication Service remain in place, even if the other trusted services, or the blockchain creation mechanism, become unavailable.

%% file: images/zipperchain.tex
\begin{tikzpicture}[node distance=1.5cm]
\node (b-1) [block] {$\block_{\idx-1}$};
\node (t-1) [block, above right = 1.5cm and 0.7cm of b-1] {$\time_{\idxj-1}$};
\node (b)   [block, below right= 1.5cm and 0.7cm of t-1] {$\block_{\idx}$};
\node (t)   [block, above right= 1.5cm and 0.7cm of b] {$\time_{\idxj}$};
\node (b+1) [block, below right= 1.5cm and 0.7cm of t] {$\block_{\idx+1}$};
\node (t+1) [block, above right= 1.5cm and 0.7cm of b+1] {$\time_{\idxj+1}$};


\node (s-1) [block, above = 1.5cm of t-1] {$\seq_{\idxk-1}$};
\node (s)   [block, above = 1.5cm of t] {$\seq_{\idxk}$};
\node (s+1) [block, above = 1.5cm of t+1] {$\seq_{\idxk+1}$};

\draw [farrow] (t-1) -- (b-1);
\draw [farrow] (b) -- (t-1);
\draw [farrow] (t) -- node[fill=backgroundcolor] {$\link(\block_{\idx})$} (b);
\draw [farrow] (b+1) -- node[fill=backgroundcolor] {$\link(\time_{\idxj})$} (t);
\draw [farrow] (t+1) -- (b+1);

\node (m-1) [database, below= 1.2cm of b-1] {$\merkle_{\idxl-1}$};
\node (m)   [database, below= 1.2cm of b, aspect=0.15] {$\merkle_{\idxl}$};
\node (m+1) [database, below= 1.2cm of b+1] {$\merkle_{\idxl+1}$};

\draw [farrow] (b-1) -- (m-1);
\draw [farrow] (b) -- node[fill=backgroundcolor] {$\link(\merkle_{\idxl})$} (m);
\draw [farrow] (b+1) -- (m+1);

\draw [farrow] (s-1) -- (t-1);
\draw [farrow] (s) -- node[fill=backgroundcolor] {$\link(\time_{\idxj})$} (t);
\draw [farrow] (s+1) -- (t+1);

\end{tikzpicture}

%% file: images/fork.tex
\begin{tikzpicture}[node distance=1.5cm]
\node (b-1) [block] {$\block_{\idx-1}$};
\node (t-1) [block, above right= 1cm and 0.25cm of b-1] {$\time_{\idxj-1}$};
\node (b)   [block, below right= 1cm and 0.25cm of t-1] {$\block_{\idx}$};
\node (t)   [block, above right= 1cm and 0.25cm of b] {$\time_{\idxj}$};
\node (b+1) [block, below right= 1cm and 0.25cm of t] {$\block_{\idx+1}$};
\node (t+1) [block, above right= 1cm and 0.25cm of b+1] {$\time_{\idxj+1}$};
\node (b+2) [block, below right= 1cm and 0.25cm of t+1] {$\block_{\idx+2}$};
\node (t+2) [block, above right= 1cm and 0.25cm of b+2] {$\time_{\idxj+2}$};

\draw [farrow] (t-1) -- (b-1);
\draw [farrow] (b) -- (t-1);
\draw [farrow] (t) -- (b);
\draw [farrow] (b+1) -- (t);
\draw [farrow] (t+1) -- (b+1);
\draw [farrow] (b+2) -- (t+1);
\draw [farrow] (t+2) -- (b+2);

\node (bp)   [block, below = 2cm of b] {$\block'_{\idx}$};
\node (tp)   [block, below right= 1cm and 0.25cm of bp] {$\time'_{\idxj}$};
\node (bp+1) [block, above right= 1cm and 0.25cm of tp] {$\block'_{\idx+1}$};
\node (tp+1) [block, below right= 1cm and 0.25cm of bp+1] {$\time'_{\idxj+1}$};

\draw [farrow] (bp) -- (t-1.south);
\draw [farrow] (tp) -- (bp);
\draw [farrow] (bp+1) -- (tp);
\draw [farrow] (tp+1) -- (bp+1);

\draw let \p1 = (b), \p2 = (bp), \p3 = (t-1) in
(\x3,\y2*0.5+\y1*0.5) node [seq_block] (s-1) [anchor=center] {$\seq_{\idxk-1}$};
\node (s) [seq_block, anchor=center, right= 0.45cm of s-1] {$\seq_{\idxk}$};
\node (s+1) [seq_block, anchor=center, right= 0.45cm of s] {$\seq_{\idxk+1}$};
\node (s+3) [seq_block, anchor=center, right= 0.45cm of s+1] {$\seq_{\idxk+2}$};
\node (s+2) [seq_block, anchor=center, right= 0.45cm of s+3] {$\seq_{\idxk+3}$};
\node (s+4) [seq_block, anchor=center, right= 0.45cm of s+2] {$\seq_{\idxk+4}$};

\draw [seq_arrow] (s-1) -- (t-1);
\draw [seq_arrow] (s) -- (t);
\draw [seq_arrow] (s+1) -- (tp);

\draw [seq_arrow] (s+2) -- (t+1);

\draw let \p1 = (b+2), \p2 = (t+2) in (\x2,\y1) node (a1) [anchor=center] {};
\draw [seq_arrow] (s+3) -- (a1.center) -- (t+2);

\draw [seq_arrow] (s+4) -- (tp+1);


\end{tikzpicture}

%% file: 4.implementation.tex
\section{Implementation}
\label{sec:implementation}

The purpose of \sol\ is to provide an ordering service over an incrementally growing set of transactions.
\sol\ brings this service to users through the following abstract interface:
\begin{equation}
  \notag
  \begin{split}
    & ~\text{\writef}(\tx) \\
    \cert \coloneqq& ~\text{\verify}(\genesis{\block}, \tx)
  \end{split}
\end{equation}

The \writef\ function takes the transaction data $\tx$ as input and starts the process of recording $\tx$ on a \sol\ blockchain.
The \writef\ function is invoked by users on cloud infrastructure in the context of a particular chain's genesis block $\genesis{\block}$.
The \verify\ function takes a \sol's genesis block $\genesis{\block}$ and the transaction $\tx$ as input to return a certificate~$\cert$. 
We design the \verify\ function to run on user infrastructure, and so it needs $\genesis{\block}$ for context.

Assume that $\tx$ is in at least one true triad on the main chain and let $\tri$ be the lowest-height main chain true triad containing the transaction.
We define a certificate 
\mbox{$\cert = \langle \tx\!\!, \genesis{\block}^\link\!\!, \ts, \heightval, \rank  \rangle$,} 
where
\mbox{$\cert.\tx = \tx$} is the transaction representation in \sol,
\mbox{$\cert.\genesis{\block}^\link = \link(\genesis{\block})$} identifies the chain,
\mbox{$\cert.\ts = \tri.\time.\ts$} is the timestamp of $\tx$,
\mbox{$\cert.\heightval = \height(\tri)$} is the height of the transaction, and
\mbox{$\cert.\rank = \indexf(\tx, \tri.\block.\merkle.\leaves)$} is the rank of the transaction in the leaves of its Merkle tree.

The transaction is \emph{finalized} when all elements of $\tri$, all \mbox{elements} of the preceding main chain triads, and all sequence attestations with counter values smaller than $\tri.\seq.\ctr$ have been replicated by the trusted \repsvclong.
For a user, obtaining a certificate over transaction $\tx$ is the observable proof of finalization of $\tx$.

When a user trusts the \verify\ function they know that:
\begin{enumerate}
    \item The transaction $\tx$ (identified as $\cert.\tx$) was finalized on chain $\cert.\genesis{\block}^\link$ and will remain unchanged on that chain
    \item The transaction $\tx$ existed before time $\cert.\ts$
    \item Given $\cert'$ over another transaction $\tx'$, the transaction $\tx$ precedes transaction $\tx'$ if and only if
    \begin{itemize}
        \item $\cert.\genesis{\block}^\link = \cert'\!\!.\genesis{\block}^\link$ and
        \item $\cert.\heightval < \cert'\!\!.\heightval ~~\lor~ (\cert.\heightval = \cert'\!\!.\heightval \land \cert.\rank < \cert'\!\!.\rank)$
    \end{itemize}
\end{enumerate}
Thus, by providing transaction certificates, \sol\ provides a transaction ordering service.

The rest of this section describes the process of recording and verifing a \sol\ transaction.
\refsec{sec:implementation:write} describes the \writef\ process that starts with a user submitting the hash of a transaction to a set of \sol\ services running in the cloud.
\refsec{sec:implementation:omc} and \refsec{sec:implementation:verify} describe the \verify\ process of the user downloading \sol\ triads to independently determine, on the user's machine, the set of main chain triads and a certificate of the transaction.

\subsection{Recording a Transaction}
\label{sec:implementation:write}

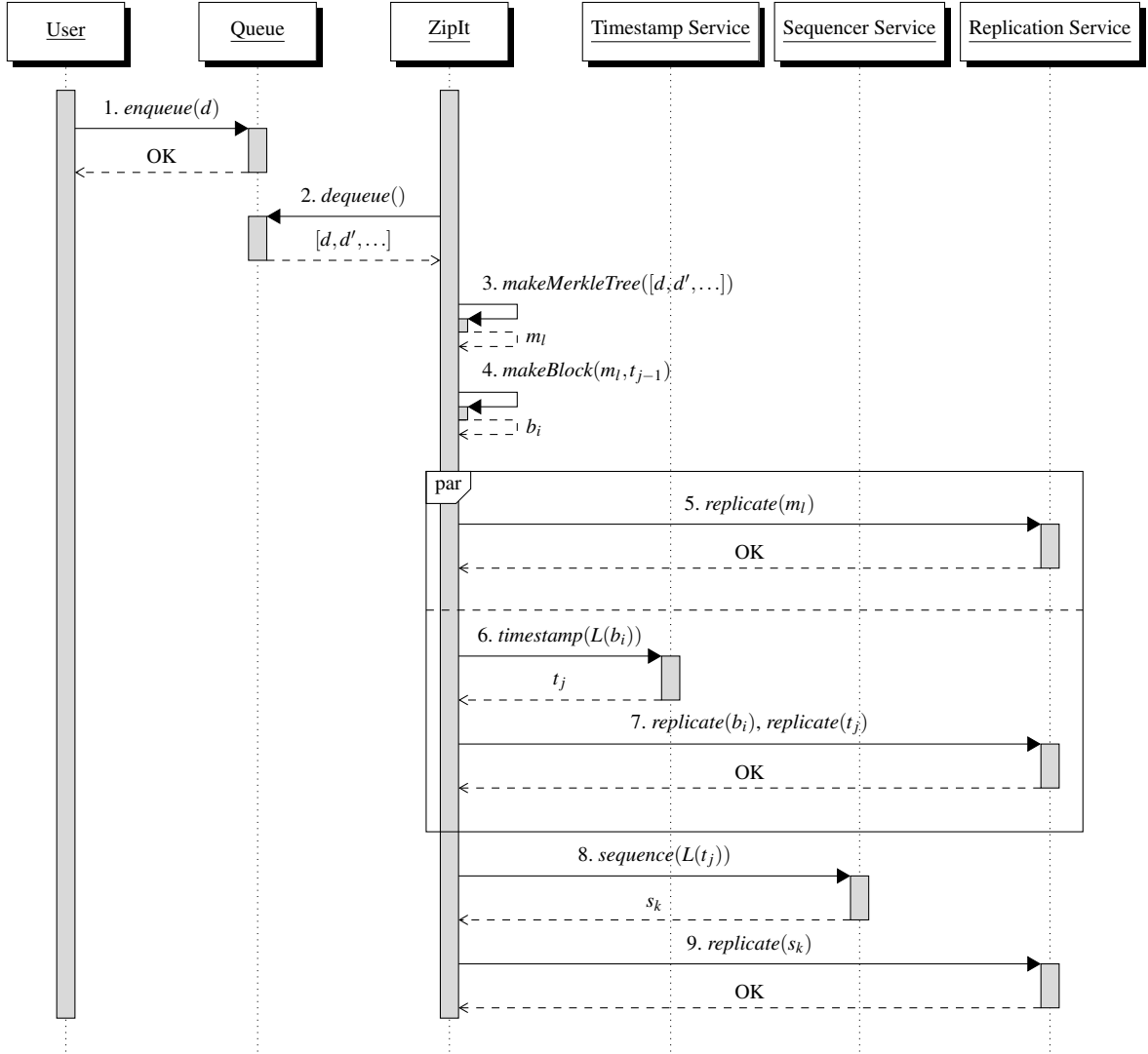
\begin{figure*}[h]
    \centering
    \input{images/write_sequence}
    \vspace{-5pt}
    \caption{Process implementing the \writef\ function.}
    \label{fig:write}
\end{figure*}

\reffig{fig:write} shows the process implementing the \writef\ function in \sol.
The numbering of the process description below corresponds to the arrow numbering in the figure.

\begin{enumerate}
\item
\label{step:write:user->queue}
To record a transaction $\tx$ on \sol, a \userlong\ calls \writef($\tx$), which invokes the $\enqueue(\tx)$ function of a cloud \queue\ service.

\item 
\label{step:write:zipit->queue}
Periodically, the \zipit\ service obtains a FIFO-ordered batch of transactions $\batch$ by invoking the \dequeue\ function of the cloud \queue\ service.

\item
\label{step:write:zipit-merkle}
\zipit\ creates a new Merkle tree by invoking an internal \merklef\ function with the batch of transactions as input.
The function returns a new Merkle tree $\merkle_\idxl$,
where
$\merkle_\idxl.\uuid$ is a newly created UUID and 
$\merkle_\idxl.\leaves = \batch$.

\item
\label{step:write:zipit-block}
\zipit\ creates a new block by invoking an internal \mbox{\blockf} function with the Merkle tree $\merkle_\idxl$ and the link $\link(\time_{\idxj-1})$ to the preceding timestamp attestation $\time_{\idxj-1}$ as input.\footnotemark
\footnotetext{Since a \sol\ chain starts by replicating the genesis triad $\genesis{\tri}$, \zipit\ always has a $\link(\time_{\idxj-1})$ to the previous timestamp attestation to include in a block.}
The function returns a block 
$\block_\idx$,
where
$\block_\idx.\uuid$ is a newly created UUID, 
$\block_\idx.\merkle^\link = \link(\merkle_\idxl)$, and
$\block_\idx.\time^\link = \link(\time_{\idxj-1})$.

\item 
\label{step:write:zipit->replicate-leaves}
\zipit\ replicates the Merkle tree $\merkle_\idxl$ by invoking the \mbox{\replicate} function of a trusted \repsvclong\ with $\merkle_\idxl$ as input.\footnotemark
\footnotetext{To conserve space, it is sufficient to replicate only the ordered leaves of the Merkle tree, since we can reconstruct the data structure $\merkle_\idxl$ from $\link(\merkle_\idxl)$ and $\merkle_\idxl.\leaves$ on retrieval.}

\item 
\label{step:write:zipit->timeit}
In parallel with step \ref{step:write:zipit->replicate-leaves},
\zipit\ creates a timestamp attestation $\time_\idxj$ over block $\block_\idx$ by invoking the \stamp\ function of a trusted \timesvclong\ with $\link(\block_\idx)$ as input.

\item 
\label{step:write:zipit->replicate-zigzag}
Still in parallel with step \ref{step:write:zipit->replicate-leaves},
\zipit\ replicates block $\block_\idx$ and timestamp attestation $\time_\idxj$.

\item 
\label{step:write:zipit->seqit}
Once the Merkle tree $\merkle_\idxl$ and the block $\block_\idx$ have been replicated, 
steps \ref{step:write:zipit->replicate-leaves} and \ref{step:write:zipit->replicate-zigzag} returned ``OK,''
it is safe to assign the timestamp attestation a sequence number.
\zipit\ creates a sequence attestation $\seq_\idxk$ by invoking the \sequence\ function of a  trusted \seqsvclong\ with $\link(\time_\idxj)$ as input.

\item 
\label{step:write:zipit->replicate-seq}
Finally, \zipit\ replicates $\seq_\idxk$ completing the $\writef$ process.\footnotemark
\footnotetext{
A sequence attestation $\seq$ does not have a UUID to use in the \linkf\ function in \replicate.
Instead, $\seq$ associates sequence numbers with a \tttt{sequenceID} represented by a UUID and signed by $\key^-_\seqsvc$ in $\seq.\sig$.
To create unique references to sequence attestations we redefine the \linkf\ function as
\mbox{$\link(\seq) = \langle \texttt{sequenceID} \mdoubleplus \seq.\ctr, \hash(\seq) \rangle$}.
}
\end{enumerate}

It is important to note that the \writef\ function does not guarantee that a transaction has been recorded on a \sol\ blockchain.
Indeed, the \writef\ function returns after step~\ref{step:write:user->queue} notifying the \userlong\ that the transaction has been accepted for processing.
The \userlong\ knows that a transaction has been recorded on the blockchain only after the \verify\ function returns a certificate.
The User can expect that \verify\ will produce the correct result on a transaction only after the transaction is finalized.

Notice that the construction method presented in \reffig{fig:write} will produce main chain triads with monotonically increasing sequence attestation counter values.
We imagine the possibility of other construction methods that relax the barrier before step \ref{step:write:zipit->seqit} and produce main chain triads with non-monotonic sequence attestation counter values, such as $\tri_{k+3}$ and $\tri_{k+2}$ in \reffig{fig:fork}.
Regardless, the algorithm to determine which triads are on the main chain, presented in the next section, does not require sequence attestation counter monotonicity.

\subsection{On Main Chain}
\label{sec:implementation:omc}

Let $\blocks, \times, \seqs$ be a set of blocks, timestamp attestations, and sequence attestations, respectively.
Recall that $\genesis{\block}$ is the genesis block (containing $\key^+_\seqsvc$ is the public key of the sequencer service and $\key^+_\timesvc$ is the public key of the timestamp service)  and $\encl$ is the well-known enclave attestation of the sequencer.
The algorithm,  \mbox{$\omc(\genesis{\block}, \encl, \blocks, \times, \seqs)$}
produces a sequence of true triads on the main chain.
We provide the details of \omc\ in \refalg{alg:omc}.

From a high level, \reflines{alg:omc:initZ}{alg:omc:endInit} initialize, verify keys, check the sequence attestation for $\genesis{\block}$ and validate the timestamp attestation for $\genesis{\block}$.
Upon completion of these lines, the main chain $\zpr$ is initialized with a true triad containing $\genesis{\block}$.

\reflines{alg:omc:beginLoop}{alg:omc:endLoop} grows $\zpr$ using the definition of the main chain.
Specifically, to find the next true triad, it is sufficient to find the lowest order block $\block^*$ whose parent is in the last true triad of $\zpr$ (along with a few additional conditions and supporting objects).
\reflines{alg:omc:loopBeginComment}{alg:omc:findMinTriad} find a candidate lowest order successor in $\blocks$, called $\minTriad.\block$, though it may not be $\block^*$.
The two may not be the same when the verification process lacks some object data, for example, due to replication delay, or download failure.
\reflines{alg:omc:collectUnseen}{alg:omc:markAsSeen}, use properties of the sequence attestations to rule out the existence of a $\block^*~\neq~\minTriad.\block$.

\begin{algorithm}
\caption{$\omc(\genesis{\block}, \encl, \blocks, \times, \seqs)$}
\label{alg:omc}
\begin{algorithmic}[1]

\State $\zpr \gets [~]$
\Comment{Main chain triads}
\label{alg:omc:initZ}

\State $\ctr \gets -1$
\Comment{Boundary of seen true triad sequence numbers}

\LComment{Load public keys from the genesis block}
\State $\key^+_\timesvc \gets \genesis{\block}.\key^+_\timesvc$
\State $\key^+_\seqsvc \gets \genesis{\block}.\key^+_\seqsvc$

\If{$\decrypt{\key_\enclsvc^+}{\encl}.\key \neq \key^+_\seqsvc$}
    \State \Return $\varnothing$ \Comment{Unattested sequencer service public key}
\EndIf

\LComment{Check block $\genesis{\block}$}
\BeginBox[draw=blue,dashed,thick]
\If{$
    \exists \block \in \blocks, \time \in \times, \seq \in \seqs ~|~
        \truetri\left( \langle \block, \time, \seq \rangle, \key^+_\timesvc, \key^+_\seqsvc \right) \land 
    \mbox{~~}\link(\block) = \link(\genesis{\block}) \land 
        \seq.\ctr = 0
$}
\label{alg:omc:checkGenesis}
\EndBox
    \State $\zpr \gets \zpr \mdoubleplus \langle \block, \time, \seq \rangle$
    \label{alg:omc:initZpr}
    \Comment{Add 0th triad to main chain}
    \State $\ctr \gets 0$
    \label{alg:omc:initCtr}
    \Comment{Mark true triad with seq. num. 0 as seen}
\Else
    \State \Return $\varnothing$
    \Comment{Block $\genesis{\block}$ was invalid}
\EndIf
\label{alg:omc:endInit}

\vspace{5pt}
\LComment{Add main chain triads to $\zpr$}
\Loop
\label{alg:omc:beginLoop}

\LComment{Find all true triad successors to last triad in $\zpr$}
\label{alg:omc:loopBeginComment}

\BeginBox[draw=PineGreen,dashdotted,thick]
\State
    $\begin{array}{@{}l@{}l}
        \succs \gets \{ \langle \block, \time, \seq \rangle ~|~
        &~\block \in \blocks \land
        \time \in \times \land
        \seq \in \seqs \land \\
        &~\truetri\left( \langle \block, \time, \seq \rangle, \key^+_\timesvc, \key^+_\seqsvc \right) \land \\
        &~\block.\time^\link = \link(\zpr.\tail.\time) 
        \}
    \end{array}$ \label{alg:omc:succs}
\EndBox
\If {$|\succs| = 0$}
    \State \Return \zpr \Comment{No more successors}
\EndIf

\LComment{Find successor with minimal sequence attestation}
\State $\minTriad \gets
    \argmin_{\succElt \in \succs} \succElt.\seq.\ctr$
    \label{alg:omc:findMinTriad}

\LComment{Fill in the sequence gap}
\If {$\ctr \le  \minTriad.\seq.\ctr $}
    \label{alg:omc:collectUnseen}

    \LComment{Find true triads between $\ctr$ and $\minTriad.\seq.\ctr$}
    \State
        $\begin{array}{@{}l@{}l}
        \gap \gets  \{ \langle \block, \time, \seq \rangle ~|~
            &~\block \in \blocks \land
            \time \in \times \land
            \seq \in \seqs \land \\
            &~\truetri\left( \langle \block, \time, \seq \rangle, \key^+_\timesvc, \key^+_\seqsvc \right) \land \\
            &~\seq.\ctr \in (\ctr, \minTriad.\seq.\ctr)
            \}
        \end{array}$ \label{alg:omc:gap}

    \If {$|\gap| \neq  \minTriad.\seq.\ctr - \ctr - 1$}
        \label{alg:omc:isGapFilled}
        \State \Return \zpr \Comment{Missing a triad}
    \EndIf

    \State $\ctr \gets \minTriad.\seq.\ctr$
        \Comment{Mark true triads up to $\minTriad$ as seen}
        \label{alg:omc:markAsSeen}
\EndIf

\State $\zpr \gets \zpr \mdoubleplus \minTriad$
    \Comment{Extend the main chain}
    \label{alg:omc:extendChain}
\EndLoop
\label{alg:omc:endLoop}

\end{algorithmic}

\begin{tikzpicture}[node distance=1.5cm]
\node (b)   [block] {$\genesis{\block}$};
\node (t)   [block, above right= 1.5cm and 0.7cm of b] {$\time$};

\node (s)   [seq_block, right= 0.7cm of b] {$\seq^*$};
\node (text) [gray, right= 0.4cm of s] {where $\seq^*\!\!\!.\ctr = 0$};

\draw [farrow] (t) -- (b);

\draw [seq_arrow] (s.north) --  (t);

\node (p1)  [left = 4.5cm of t.north, anchor=center] {};
\node (p2)  [left = 4.5cm of s.south, anchor=center] {};
\draw [blue, dashed, thick] (p1.center) -- (p2.center);
\end{tikzpicture}

\vspace{10pt}

\begin{tikzpicture}[node distance=1.5cm]
\node (t-1) [block] {$\zpr.\tail$};
\node (b)   [block, below right= 1.5cm and 0.7cm of t-1] {$\block$};
\node (t)   [block, above right= 1.5cm and 0.7cm of b] {$\time$};

\node (s)   [seq_block, right= 0.7cm of b] {$\seq$};

\draw [farrow] (b) -- (t-1);
\draw [farrow] (t) -- (b);

\draw [seq_arrow] (s.north) --  (t);

\node (p1)  [left = 5.5cm of t.north, anchor=center] {};
\node (p2)  [left = 5.5cm of s.south, anchor=center] {};
\draw [PineGreen, dashdotted, thick] (p1.center) -- (p2.center);
\end{tikzpicture}

\end{algorithm}

\begin{algorithm}
\caption{$\makecert(\tx, \merkles, \zpr)$}
\label{alg:makecert}
\begin{algorithmic}[1]
\LComment{Find the first block whose Merkle tree contains txn $\tx$}
\For{$ \tri \in \zpr$}
\Comment{Ordered iteration}
    \If{$\exists \merkle \in \merkles ~|~
    \tx \in \merkle.\leaves \land 
    \tri.\block.\merkle^\link = \link(\merkle.\leaves)
    $}
        \LComment{Found $\tx$ in a block. Return a certificate.}
        \State \Return $\langle 
        \tx,
        \link(\zpr\texttt{[}0\texttt{]}.\block), 
        \tri.\time.\ts,
        \height(\tri.\block), \\
        \mbox{~~~~~~~~~~~~~~} \indexf(\tx, \merkle.\leaves)
        \rangle$
    \EndIf
\EndFor

\State \Return $\varnothing$
\Comment{No main chain block contains transaction $\tx$}
\end{algorithmic}
\end{algorithm}

\subsection{Verifying a Transaction}
\label{sec:implementation:verify}

After calling the \writef\ function on a transaction, the \userlong\ needs to verify that the transaction was written onto a \sol\ blockchain, i.e. that the transaction exists in a finalized block on the main chain.

The verification process proceeds as follows:
\begin{enumerate}[1.]
\item
To verify a transaction $\tx$ a \userlong\ calls
$\verify(\genesis{\block}, \tx)$ by passing in 
the genesis block $\genesis{\block}$ and transaction $\tx$.
As mentioned earlier, the \verify\ function can execute on the \userlong's machine, and so the \userlong\ does not need to trust BLOCKY to execute the function correctly.

\item
\label{step:verify:download}
The \verify\ function downloads the current state of a \sol\ blockchain from the \repsvclong.
Specifically, this state comprises the set of blocks~($\blocks$), Merkle trees~($\merkles$), timestamp attestations~($\times$), and sequence attestations~($\seqs$).
In practice, these sets and their verification as described below may be cached~(bootstrapped) and extended as a \sol\ chain grows.

\item
Next, the \verify\ function determines which blocks and timestamp attestations are on the main chain.
To do so, \verify\ calls the function \omc, shown in \refalg{alg:omc}, which produces a set of alternating main chain blocks and timestamp attestations
\mbox{$\zpr = \omc(\genesis{\block}, \encl, \blocks, \times, \seqs)$}.
Recall that $\encl$ is the enclave attestation over the public key of the sequencer service signed with the well-known public key $\key_\enclsvc^+$.

\item
Finally, the \verify\ function determines whether any main chain block contains the hash of transaction~$\tx$.
We assume that transactions are idempotent and their hashes unique, and so the certificate of a transaction always pertains to its first instance.
To create a certificate, \verify\ calls the \mbox{\makecert} function, shown in \refalg{alg:makecert}, which produces a certificate
$\cert = \makecert(\tx, \merkles, \zpr)$.
If $\cert \neq \varnothing$, \mbox{\verify} returns the certificate to the \userlong.
\end{enumerate}

When the \verify\ function returns a certificate $\cert$, the function asserts that the transaction  $\tx$ existed at time $\cert.\time.\ts$ on the \sol\ ledger $\genesis{\block}.\uuid$.
When \verify\ returns a certificate it also indicates that the transaction is finalized and will not change on the \sol\ ledger.



%% file: images/write_sequence.tex
    \begin{sequencediagram}
        \tikzstyle{inststyle}+=[top color=backgroundcolor, bottom color=backgroundcolor]
        \newthread{user}{\userlong}
        \newinst[1]{queue}{\queue}
        \newthreadShift{zipit}{\zipit}{1cm}
        \newinst[1]{timeit}{\timesvclong}
        \newinst{seqit}{\seqsvclong}
        \newinst{repit}{\repsvclong}

        \begin{call}
            {user}{\refstep{step:write:user->queue} $\enqueue(\tx)$}
            {queue}{OK}
        \end{call}

        \begin{call}
            {zipit}{\refstep{step:write:zipit->queue} $\dequeue()$}
            {queue}{$\batch$}
        \end{call}
        
        \begin{callself}
            {zipit}{\refstep{step:write:zipit-merkle} $\merklef(\batch)$}
            {$\merkle_\idxl$}
        \end{callself}

        \begin{callself}
            {zipit}{\refstep{step:write:zipit-block} $\blockf(\merkle_\idxl, \time_{\idxj-1})$}
            {$\block_\idx$}
        \end{callself}

        \begin{sdblock}{par}{}
            \addtocounter{seqlevel}{-1}
            
            \begin{call}
                {zipit}{\refstep{step:write:zipit->replicate-leaves} $\replicate(\merkle_\idxl)$}
                {repit}{OK}
            \end{call}

            \stepcounter{seqlevel} 
            
            \begin{call}
                {zipit}{\refstep{step:write:zipit->timeit} $\stamp(\link(\block_\idx))$}
                {timeit}{$\time_\idxj$}
            \end{call}
            \begin{call}
                {zipit}{\refstep{step:write:zipit->replicate-zigzag} $\replicate(\block_\idx)$, $\replicate(\time_\idxj)$}
                {repit}{OK}
            \end{call}

        \end{sdblock}
        \path (nw)--coordinate (midpoint) (se); 
        \node (shifted) at ([yshift=0.6cm]midpoint) {}; 
        \draw[dashed] (nw|-shifted)--(shifted-|se);
            
        \begin{call}
            {zipit}{\refstep{step:write:zipit->seqit} $\sequence(\link(\time_\idxj))$}
            {seqit}{$\seq_\idxk$}
        \end{call}
        \begin{call}
            {zipit}{\refstep{step:write:zipit->replicate-seq} $\replicate(\seq_\idxk)$}
            {repit}{OK}
        \end{call}
        
    \end{sequencediagram}

%% file: 5.evaluation.tex
\section{Evaluation}
\label{sec:evaluation}

To understand the performance of \sol\ we load-tested it as follows.
Referring to \reffig{fig:write}, we deployed the \queue\ service, implemented on a Redis database, and the \zipit\ service on a single \ttt{t2.xlarge} instance.
We connected \zipit\ to a \timesvclong\ implemented on AWS Cognito, a \seqsvclong\ running in a Nitro Enclave on a \ttt{c5a.xlarge} node, and a \repsvclong\ running a two-out-of-three replication scheme on two AWS buckets and one Azure bucket each in a different availability zone.
Finally, we simulated traffic from \userlong s by running Graphana's k6 load testing tool on a \ttt{t2.2xlarge} node~\cite{Graphana25k6}.

We performed two experiments. 
First, under the `no load' scenario, we sent nominal traffic to measure block finality, or the time difference between the $\enqueue(\tx)$ message and the OK message following the $\replicate(\seq_\idxk)$ call.
Second, under the `load' scenario, we simulated traffic from 250 \userlong s to measure block finality and transaction throughput measured as the number of OK messages following the $\enqueue(\tx)$ call.
Notice that there is no back-pressure to k6 based on the number of transactions inserted into the queue.
Each $\dequeue()$ call collects all the messages in the queue and we ran the \omc\ algorithm on a separate machine to ensure the transactions issued by k6 do indeed get recorded in the replicated \sol\ data structures.

\begin{figure}[t]
    \centering
    \includegraphics[width=\columnwidth]{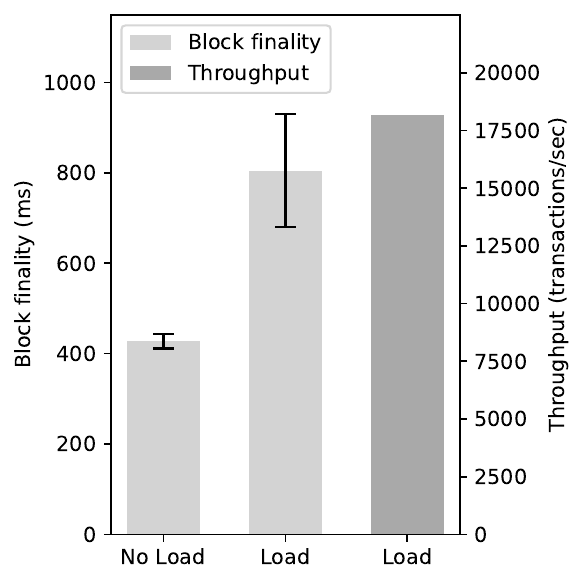}
    \caption{\sol\ block finality and throughput.}
    \label{fig:evaluation}
\end{figure}

\reffig{fig:evaluation} shows the performance of \sol\ under the no load and load scenarios.
The left x-axis marks block finality in milliseconds and the right y-axis marks throughput in transactions per second.
The block finality bars show mean values with 95\% confidence intervals.

We observe that the no load scenario achieves mean block finality of 427\,ms, while under load block finality increases to 805\,ms on average with the throughput of 18,183 transactions per second.
In the load scenario we limit the the traffic to 250 users, which we determined experimentally to reliably keep the p90 of block latency under 1\,s.
The increase in finality is due the $\replicate(\merkle_\idxl)$ step taking longer to encode transaction hashes as shards, ensure they are linearly independent, and replicate them.
Note that the results in \reffig{fig:evaluation} are for a fairly unoptimized version of \sol\ (e.g. we send all messages serialized as JSON), and so should be treated as preliminary.

%% file: 6.discussion.tex
\section{Discussion}
\label{sec:discussion}

\sol\ guarantees immutability, agreement, and availability.
The design of \sol\ is quite different from that of other blockchains and the implications of these differences are worth discussing.

\vspace{-5pt}
\paragraph{Tokens}

\sol, by design, does not have a native token.
The continued operation of a \sol\ instance depends on payments to cloud service providers to operate the \sol\ construction process and the third-party services on which it relies.
As such, \sol\ is gas-less and does not face regulator challenges of coin-emitting chains~\cite{WhiteHouse22Responsible}.
At the same time, a smart contract system supported by \sol\ could interface with a diverse set of payment rails, including established crypto or fiat currencies, as required by the distributed applications it supports.

\vspace{-5pt}
\paragraph{Smart Contracts}

At the moment \sol\ does not implement a smart contract functionality.
However, since \sol\ provides a total order of transactions, if a transaction were the publication, or an invocation of a smart contract, that transaction could be executed deterministically off-chain based on the resulting state of the preceding transactions.
Offchain transaction execution has already been successfully demonstrated by several blockchain implementations~\cite{Hentschel19Flow, Cheng19Ekiden, Zyskind16Enigma}.
We prototyped a similar approach forwarding transactions recorded on \sol\ to an EVM for execution.
Differently from these blockchains, however, we do not plan on creating a native \sol\ token.
Instead, we plan to support cryptocurrency operations based on coins transferred to a \sol\ smart contract using a cross-chain approach~\cite{Kwon22Cosmos, Wood16Polkadot}.
For more general compute, it would also be possible to order invocations to the BLOCKY Attestation Service as well as the resulting attestations to maintain canonical application state~\cite{Block25AS}.


\vspace{-5pt}
\paragraph{Interoperability}

Since \sol\ does not require a community of validators for its security, it is possible for many \sol\ ledgers to coexist without weakening their correctness guarantees.
To create a \sol\ ledger, a developer needs to stand up the infrastructure to implement the \writef\ process described in \refsec{sec:implementation:write}, which makes it practical for each distributed application to use its own \sol\ app chain.
With the introduction of a smart contract system on top of \sol\ it will be possible for one \sol\ app chain to verify the certificates of transactions from another, which will simplify bridging between smart contracts running on different \sol\ app chain.
Note, however, that a total order of transactions is maintained within a \sol\ ledger, but not across them.

\vspace{-5pt}
\paragraph{Service Interactions}

An inquisitive reader might ask why do we need both a \timesvclong\ and a \seqsvclong\ to support blockchain correctness guarantees.
We see the cooperation between these services as essential.
A \timesvclong, as implemented by Amazon Cognito, Auth0, or other OAuth services, provides trustworthy real-world timestamps, but cannot provide order, since physical timestamps in two timestamp attestations can be equal and since a timestamp attestation can always be withheld and revealed at a later time.
On the other hand a \seqsvclong, as implemented by Amazon Nitro Enclaves or possible on hardware-based TEEs, allows users to alter the system clock.
To have both, real-world timestamps and order both services are needed.

It may be possible for Nitro Enclaves, or another TEE system, to provide trustworthy physical timestamp in the attestations.
In that case one could think that it is possible to create a blockchain on the enclave itself.
The problem with that approach is that it does not allow a TEE failure recovery mechanism without horizontal scaling, which creates the possibility of forks.

\vspace{-5pt}
\paragraph{Trust model}

Blockchains base their correctness guarantees on the correctness of blockchain algorithms, the correctness of the implementation of these algorithms, and on a set of assumptions that constrain the power of a determined attacker.
The correctness guarantees of a \pow\ blockchain, such as Bitcoin, comes from the correctness of the Nakamoto consensus algorithm~\cite{Nakamoto08Bitcoin}, the verifiable open-source implementation of the Bitcoin protocol~\cite{22BitcoinCore}, and on the assumption that 51\% of the hash power is controlled by honest nodes, who follow the Bitcoin protocol.
On the other hand \pos\ blockchains rely on the assumption that two-thirds of value staked by nodes remains honest.
These assumptions may not always be sufficient, for example Eyal\etal~\cite{Eyal18Majority} have shown that you need less than 50\% to attack Bitcoin, or hold true, for example when a small number of validators allows capture~\cite{Uno.Reinsure22Biggest}.

Similarly to other blockchains, \sol's mechanisms and implementation will be publicly available for inspection.
Where \sol\ differs is in the set of assumptions behind its correctness guarantees.
Specifically, we assume that an attacker cannot gain control of the trusted \timesvclong, \seqsvclong, and \repsvclong.
We believe this assumption is reasonable and acceptable to many users, since our implementations of these services are based on third-party services from major cloud providers, and so are backed by their security teams.
The capture of a user authentication service service would lead to significant reputational and financial loss for its operator, which creates strong incentives for Amazon as well as Auth0 to prevent such a scenario.
The same is true for the other third-party services on which \sol\ relies.

The correctness guarantees of \sol\ continue to hold even in the face of the failure of any one these services.
To create a fork on a \sol\ ledger, or violate the agreement guarantee, an attacker would need to create an alternative history of valid triads.
To do that an attacker would need to gain control of the user authentication service and Nitro Enclave private keys and to break the Amazon instance isolation.
To alter the history of a \sol\ ledger, or violate the immutability and availability guarantees, an attacker would additionally need to erase an existing triad history by gaining control of at least two cloud storage services.
We think that such simultaneous failures are extremely unlikely and that the strength of \sol\ correctness guarantees is satisfactory for many blockchain applications.
While impossible to measure, we conjecture that in practice it may be more difficult take over \sol\ than other blockchain deployments backed by relatively small verifier communities.
Finally, a threat to availability comes from BLOCKY losing control of our administrative cloud service keys, which could result in inadvertent, or malicious deletion of our cloud service accounts.
Such a move would result in the loss, though not alteration, of data in cloud storage services, even if it is placed in compliance mode.
Interested parties who want to preclude this eventually may choose to back up \sol\ objects in their own deployment of the \repsvclong.

\vspace{-5pt}
\paragraph{Censorship resistance}

One property often attributed to blockchains is that of censorship resistance.
Since the implementation of \sol\ is centralized, applications based on it may be seen as vulnerable to censorship.
An operator of \sol\ may wish to block certain transactions based on content, or origin.
Such actions can be easily circumvented by submitting transactions through a system like Tor, which obfuscates sources, or through commitment schemes, that obfuscate content.
Indeed, such a combined approach has been adopted by BloXroute~\cite{Klarman18bloXroute}.

\vspace{-5pt}
\paragraph{Availability and Liveness}

Another attack could be for the operator of \sol\ to simply shut down the infrastructure to make the system unavailable.
It is important to note that such a scenario would only prevent the creation of new blocks (liveness, service availability), because existing blocks and transactions can be obtained from the \repsvclong, or from data backed up from the \repsvclong\ (data availability).

To maintain liveness then \sol\ infrastructure becomes unavailable, an interested party can identify the last block on the main chain and bring up their own \sol\ infrastructure to continue from that block.
Ultimately, it is the \seqsvclong\ that determines which blocks are on the main chain, so even if the original \sol\ infrastructure becomes available again, or interested parties bring up multiple instances of \sol\ infrastructure, only one of the blocks of the same height they produce will be able to extend the main chain.
The above approach assumes that the \seqsvclong\ continues to operate.
As mentioned in \refsec{sec:services:reliability}, we have a method to replace failing, or halted, instances of a \seqsvclong, which we will detail in another paper.
\looseness-1

%% file: 7.related_work.tex
\section{Related Work}
\label{sec:related_work}


Early blockchain designs, such as Bitcoin~\cite{Nakamoto08Bitcoin}, were made possible by a Proof-of-Work~(\pow) leading to the Nakamoto probabilistic consensus.
A miner creates a new block by solving a cryptographic puzzle and guessing a \emph{nonce}, the hash of which, together with other parts of the blockchain, produces a hash that is sufficiently small, when considered as a binary number.
While this mechanism has proven resilient to coordinated attacks, it is costly in terms of electricity used by mining hardware.
To address the cost of mining of new blocks, Peercoin~\cite{King2012PPCoin} was the first to adopt Proof-of-Stake~(\pos), where the opportunity to create the next block is decided by a lottery weighted by the number of coins staked by a verifier node, rather than the node's hash power.
The correctness of the lottery is usually enforced by a Byzantine Fault Tolerant~(\bft) consensus mechanism.
Both PoW and PoS designs have led to a number of well-established public blockchains~\cite{Nakamoto08Bitcoin,2018EOS}.

The limiting factor to the performance of these blockchains is the network performance between its miner and verifier nodes~\cite{Klarman18bloXroute}.
It simply takes some time to disseminate a new block, so that the verifiers can create its successor, rather than a fork.
The block interval then is governed by the size of the block, block interval, and network performance.
While one might naturally worry about block processing time, for example in face of complex smart contracts, verifier processing speed has not been the limiting factor to blockchain performance as of yet~\cite{Hentschel19Flow, Klarman18bloXroute}.

To gain higher performance, DLT design departs in two directions.
The first of these aims, primarily, to reduce transaction delay by decreasing block interval by using smaller blocks that take less time to disseminate.
The second aims, primarily, to increase transaction throughput by relying on a constrained number of well-connected verifier nodes.
Some blockchains combine the two approaches~\cite{22Nano, Rocket19Avalanche}.

When decreasing block interval ad~infinitum blocks become so small as to contain only a single, or a few transaction, but reference more than one previous blocks to form a directed acyclic graph~(\dag)~\cite{Rocket19Avalanche, 22Nano, Yakovenko22Solana, Gagol2019Aleph}.
In a DAG blocks at the same height in the DLT do not necessarily create a fork, which allows nodes to create them independently and tie forks together with subsequent blocks.
Transaction finality still depends on agreement among some quorum of nodes, but nodes reach agreement independently on each transaction, which tends to speed up transaction delay, but not necessarily throughput.
A good example is Nano with the impressive transaction delay of \tnu{0.146}{s}, but a relatively measly throughput of \tnu{1,\!455}{TPS}~\cite{20NanoTPS}.
\looseness-1


The second direction, constraining the number of verifiers, improves the speed of block dissemination through high-capacity, direct connections between verifiers.
A DLT may identify these verifiers through delegation as in Nano~\cite{22Nano}, or by configuring their identities into the protocol in a scheme dubbed proof-of-authority~(\poa) as in the rollout of Polkadot~\cite{22PolkadotLaunch}.
In addition to offering higher throughput and lower latency between nodes, PoA networks consume less power than PoW, and are potentially more secure than PoS since verifier identities are known~\cite{Bogdanov21PoA}.
On the other hand, the small number of nodes makes PoA networks vulnerable to 51\% and distributed denial of service~(DDoS) attacks, and their users to censorship and blacklisting~\cite{Binance20PoA}.
Not following the protocol forfeits validator reputation and with it, the right to make future blocks.  
Unfortunately, in most systems, it is not clear how to quantify reputation.

%% file: 8.conclusions.tex
\vspace{-5pt}

\section{Conclusions}
\label{sec:conclusions}
\vspace{-5pt}

BLOCKY \sol\ guarantees immutability, agreement, and availability of transaction data, but without relying on distributed consensus.
Instead, its construction process transfers trust from widely-used, third-party services onto \sol's correctness guarantees.
\sol\ runs on a pipeline of specialized services connected by a fast data center network, which allows it to 
achieve projected throughput approaching line speeds, 
reach block finality of around 100\,ms,
and operation without a native token to incentivize a community of verifiers.

%% file: 9.acknowledgments.tex
\section*{Acknowledgments}

The authors acknowledge that parts of the methodology described in this paper are covered by U.S. Patent No.~US20230179435A1~\cite{Heinecke22Patent}.

%% file: 10.appendix.tex
\onecolumn
\appendix
\section{Appendices}
\label{sec:appendix}

\input{appendices/omc}

\input{appendices/replacement_ring}

%% file: appendices/omc.tex
\subsection{On Main Chain Correctness}
Next, we prove the correctness of the $\omc$ algorithm.

\begin{theorem}
    Given,
      block $\block_0$,
      sequencer service public key $\key^+_\seqsvc$,
      timestamp service public key $\key^+_\timesvc$,
      enclave attestation $\encl$,
      a set of blocks $\blocks$,
      a set of timestamp attestations $\times$,
      and a set of sequence attestations $\seqs$,
    the algorithm $\omc(\block_0, \key^+_\seqsvc, \key^+_\timesvc, \encl, \blocks, \times, \seqs)$,
    presented in \refalg{alg:omc},
    produces a length $k$ ordered list of the first $k$ true triads of $\sol$.
\end{theorem}

\begin{proof}

    Recall that for a true triad $\tri$ to be on the main chain, its predecessor $\tri'$ must be on the main chain.
    Moreover, $\tri$ must be the lowest order successor of $\tri'$.
    Our algorithm produces the main chain iteratively by finding the lowest order child of the current tail.
    The challenge, however, is in determining if such a triad satisfies the criterion given incomplete or missing data.

    To demonstrate the correctness of our algorithm, we use the three invariants in the loop on \reflines{alg:omc:beginLoop}{alg:omc:endLoop}.
    At the beginning of iteration $\idxj$:
    \begin{enumerate}
        \item
            $\tri_0 = \zpr[0]$ is a true triad with $b_0$ and sequence number 0.
            \label{omc:loopinvar:genesis}
        \item
            Let $\ctr = \max_{\tri \in \zpr} \tri.\seq.\ctr$.
            For $\idx \in \{0, 1, \ldots, \ctr\}$, there exists a true triad $\tri$ with sequence number $\tri.\seq.\ctr = \idx$ such that $\tri$ is marked as seen.
            \label{omc:loopinvar:complete}
        \item
            $\zpr$ contains $\idxj$ true triads such that for each $\tri_\idx \in \zpr \setminus \{ \tri_0 \}$:
            \begin{enumerate}
                \item $\zpr[\idx-1] = pred(\tri_\idx)$
                \label{omc:loopinvar:omcprop:pred}
                \item for all successors of $\zpr[\idx-1]$, $\tri_\idx$ contains the minimal sequence attestation
                \label{omc:loopinvar:omcprop:min}
            \end{enumerate}
            \label{omc:loopinvar:omcprop}
    \end{enumerate}
    Informally, \refinvar{omc:loopinvar:genesis} tells us that we have a chain that started with a valid genesis block.
    \refinvar{omc:loopinvar:complete} helps us show that when we add triad $\tri$ to the main chain at height $\idxj$, it is the lowest order triad with predecessor $\tri_{\idxj-1}$, even with incomplete data.
    \refinvar{omc:loopinvar:omcprop} tells us that the triads on the main chain are valid, link back to the genesis block, and satisfy the ``on main chain rule.''
    Note that we do not keep an explicit list of ``seen'' triads as it is sufficient to store $k$.

    Consider the base case where iteration $\idxj = 1$.
    In \reflines{alg:omc:initZ}{alg:omc:endInit} we initialize.
    \refinvar{omc:loopinvar:genesis}, the predicate on \refline{alg:omc:checkGenesis} ensures that $\tri_0$ is a true triad and \refline{alg:omc:initZpr} initializes $\zpr = \lbrack \tri_0 \rbrack$.
    \refinvar{omc:loopinvar:complete}, on \refline{alg:omc:initCtr}, we initialized $\ctr = 0$ and mark $\tri_0$ as seen.
    As $\tri_0.\seq.\ctr = 0$, \refinvar{omc:loopinvar:complete} is satisfied.
    \refinvar{omc:loopinvar:omcprop}, since $\zpr$ contains just $\tri_0$, $\zpr$ contains one true triad.
    \refinvar{omc:loopinvar:omcprop:pred} and \ref{omc:loopinvar:omcprop:min} are vacuously true since the $\zpr \setminus \{\tri_0\}$ is empty.

    Let $\idxj > 1$ and assume that the invariants are true at the beginning of iteration $\idxj$ and consider the invariants at the beginning of iteration $\idxj+1$.
    Let $\zpr_\idxj = \lbrack \tri_0, \tri_1, \ldots, \tri_{\idxj-1} \rbrack$ and $\ctr_\idxj$ be the values of $\zpr$ and $\ctr$ at the start of iteration $\idxj$, respectively.
    On \refline{alg:omc:findMinTriad}, we identify a candidate $\minTriad$ and consider two cases from the branch on \refline{alg:omc:collectUnseen}.
    In the first case, $\minTriad.\seq.\ctr \le \ctr_\idxj$, and so we evaluate \reflines{alg:omc:collectUnseen}{alg:omc:markAsSeen}.
    In the second case, $\minTriad.\seq.\ctr > \ctr_\idxj$ and we can omit those steps.
    Next we consider each invariant and show the for both cases, the invariant is restored at the conclusion of the loop.

    For \refinvar{omc:loopinvar:genesis}, in both cases, the invariant remains true because we do not update $\zpr[0]$.

    For \refinvar{omc:loopinvar:complete}, in the first case, we do not evaluate \reflines{alg:omc:collectUnseen}{alg:omc:markAsSeen} and just add $\minTriad$ to $\zpr_\idxj$.
    Since $\ctr_\idxj = \max_{\tri \in \zpr_\idxj} \tri.\seq.\ctr$ (by assumption) and $\minTriad.\seq.\ctr \le \ctr_\idxj$ (branch condition), \mbox{$\ctr_\idxj = \max_{\tri \in \zpr_{\idxj} \mdoubleplus \minTriad} \tri.\seq.\ctr$}.
    Moreover, as $\minTriad.\seq.\ctr < \ctr_\idxj$ and no true triads are ever unseen, the same set of true triads that fulfil \refinvar{omc:loopinvar:complete} at the start of iteration $\idxj$ satisfy the invariant at the beginning of iteration $\idxj+1$.
    For \refinvar{omc:loopinvar:complete}, in the second case, we check that the data we have access to is complete enough such that any additional data about the chain would not produce a successor for $\tri_{\idxj-1}$ with a sequence attestation smaller than that of $\minTriad$.
    Specifically, observe that it is sufficient to show that for each $\idx \in \{\ctr_\idxj + 1, \ctr_\idxj + 2, \ldots, \minTriad.\seq.\ctr - 1 \}$, the data $\blocks, \times, \seqs$ contains a true triad $\tri$ such that $\tri.\seq.\ctr = \idx$.
    In \refline{alg:omc:gap}, we produce such a set $\gap$ of all true triads with sequence numbers in the open interval $(\ctr_\idxj, \minTriad.\seq.\ctr)$.
    As the sequence numbers are unique, to determine if any true triads are missing, in \refline{alg:omc:isGapFilled}, it is sufficient to check the size of $\gap$ (and terminate if we identify missing data).

    For \refinvar{omc:loopinvar:omcprop}, let $\zpr_{\idxj+1} = \zpr_\idxj \mdoubleplus \minTriad$, as in \refline{alg:omc:extendChain}.
    By assumption, $\zpr_\idxj$ contains $\idxj$ true triads.
    As $\minTriad \in \succs$ and all elements of $\succs$ are true triads, adding $\minTriad$ to $\zpr_\idxj$ produces $\idxj + 1$ true triads.
    For \refinvar{omc:loopinvar:omcprop:pred}, it is sufficient to show that $pred(\minTriad)$ is the tail of $\zpr_\idxj$.
    But observe that $\minTriad \in \succs$ and every element in $\succs$ is a successor of the tail of $\zpr_l$.
    For \refinvar{omc:loopinvar:omcprop:min}, it is sufficient to show that $\minTriad$ is the minimal successor of tail of $\zpr_\idxj$.
    Assume for contradiction that there exists a $\minTriad^{\!*}$ that different from $\minTriad$ with predecessor $pred(\minTriad)$ and smaller sequence attestation.
    That implies $\minTriad^{\!*}\!\!\!.\seq.\ctr < \minTriad.\seq.\ctr \le \ctr_\idxj$, however, by \refinvar{omc:loopinvar:complete} we have seen the $\ctr_\idxj + 1$ triads with sequence numbers $\{0, 1, \ldots, \ctr_\idxj\}$.
    Thus, we marked $\minTriad^{\!*}$ as seen in some previous iteration, which implies that $\minTriad^{\!*} \in \succs$ on \refline{alg:omc:succs}.
    That, however, contradicts \refline{alg:omc:findMinTriad}, which identified $\minTriad$ as the element of $\succs$ with the smallest sequence attestation.
    Therefore, \refinvar{omc:loopinvar:omcprop:min} is satisfied.

\end{proof}

%% file: appendices/replacement_ring.tex
\newpage
\subsection{Sequencer Replacement}

As mentioned in \refsec{sec:services:reliability}, \sol\ is be able generate and finalize blocks as long as the \seqsvclong\ is available.
To continue block generation, even if a \seqsvclong\ fails, we propose a \seqsvclong\ replacement mechanism.

We define a ring $\ring$ of \seqsvclong s of size $\nvar$ as a set of enclave attestations 
$\ring = \left\{ \encl_0, \encl_1, ..., \encl_{\nvar-1} \right\}$.
Since each enclave attestations $\encl$ identifies a \seqsvclong\ $\seqsvc$ by its public key $\encl.\key = \key_\seqsvc^+$, we can also think of the ring as 
$\ring = \left\{ \seqsvc^{\encl_0.\key},  \seqsvc^{\encl_1.\key}, ..., \seqsvc^{\encl_{\nvar-1}.\key} \right\}$,
where we use the superscript as the id.
For simplicity, we will denote to the ring as 
$\ring = \left\{ \seqsvc^0,  \seqsvc^1, ..., \seqsvc^{\nvar-1} \right\}$.
We require that $|\ring|$ be odd so that $|\ring| / 2 + 1$ \seqsvclong s in $\ring$ can form a majority.
Finally, we define $R$ for a particular \sol\ ledger during its creation, so that we can record it as a transaction in $\block_0$.

We modify the implementation of the \writef\ function to incorporate multiple \seqsvclong s.
\reffig{fig:write_ring} shows the updated \writef\ process.
We introduce the \sequenceit\ process with the knowledge of the state of $\ring$ and the ability to issue sequencing requests its members.

\begin{figure}[h]
\centering
\input{images/write_ring}
\vspace{-5pt}
\caption{Process implementing the \writef\ function with \seqsvclong\ replacement.}
\label{fig:write_ring}
\end{figure}

\noindent
The write process proceeds as before, but with the following modifications:
\begin{enumerate}[1.]
\setcounter{enumi}{6}

\item
\label{step:write_ring:storeit->seq}
Once a block has been replicated it is safe to assign it a sequence number.
\storeit\ sends asynchronously the hash of the timestamp attestation $\hash(\time)$
concatenated with the unique ID of the timestamp attestation $\time.\uuid$
to \sequenceit.

Asynchronously, \storeit\ also sends the UUID of the Merkle tree $\merkle.\uuid$ to \batchit\ to notify it that a block corresponding to a batch of requests has been replicated, which allows \batchit\ to stop proposing the batch to \zipit.

\item
\label{step:write_ring:seq->ring}
\sequenceit\ forwards $\langle \hash(\time), \time.\uuid \rangle$ to a \seqsvclong\ in $\ring$.

\item
\label{step:write_ring:ring->seq}
The contacted \seqsvclong\ returns a sequence attestation $\seq = \langle \bytes, \ctr, \sig \rangle$, 
where $\bytes = \langle \hash(\time), \time.\uuid \rangle$, 
and 
\mbox{$\sig = \encrypt{\key_\seqsvc^-}{\hash( \hash(\time) \mdoubleplus \time.\uuid, \,\,\ctr)}$}.

\item
\label{step:write_ring:seq->repl}
\sequenceit\ sends $\seq$ to the \repsvclong\ for replication.

\item
\label{step:write_ring:repl->zipit}
Once the \repsvclong\ confirms replication, \sequenceit\ also sends $\seq$ to \zipit\ for inclusion in a later block.

\end{enumerate}

To accommodate the inclusion of a $\seq$ into a block, we extend the definition of a block $\block$ to 
$\langle \merkle^\hash, \merkle^\uuid, \time^\hash, \time^\uuid, \uuid, \seqs \rangle$, where
$\block.\merkle^\hash$ is the the root of a Merkle tree,
$\block.\merkle^\uuid$ is its UUID,
$\block.\time^\hash$ is the hash of a~(preceding) timestamp attestation,
$\block.\time^\uuid$ is its UUID, 
$\block.\uuid$ is a distinct UUID assigned during block creation, and
$\block.\seqs$ is a set sequence attestation signatures.

Assuming $|\ring| = 3$ the structure of a \sol\ ledger might look like the one in \reffig{fig:zpr_ring_no_failures}.
We define $\seq^r_\idxk$ to be the $\idxk^\text{th}$ sequence attestations from $r^\text{th}$ sequencer in $\ring$.
And so, $\seq^0_0$ sequences $\time_0$ and is acknowledged by $\block_1$.
Notice that $\block_3$ acknowledges both $\seq^2_0$ and $\seq^1_0$, where 
\mbox{$\block_3.\seqs = \left\{ \seq^2_0.\sig, \seq^1_0.\sig \right\}$,} 
which may happen if $\seq^1_0$ is not known to \zipit\ before it makes $\block_2$.
Also notice that $\seq^0_1$ and $\seq^0_2$ both sequence $\time_3$. 
We do not want to rely on \sequenceit\ for any load balancing guarantees, and so it may be possible, though not efficient, that many sequencers sequence the same timestamp attestation.

\begin{figure}[h]
\centering
\vspace{-10pt}
\input{images/zpr_ring_no_failures}
\vspace{-5pt}
\caption{\sol\ with replacement ring sequence attestations.}
\label{fig:zpr_ring_no_failures}
\end{figure}

Let us now consider scenarios, in which a sequence attestation is missing either because \seqsvclong\ fails, or because it sequences a timestamp attestation that is not on the main chain.
\reffig{fig:zpr_ring_with_fork} shows a fork, in which both $\block_2$ and $\block'_2$ include $\time_1$.
In \reffig{fig:zpr_ring_with_fork}, $\seq^2_0$ sequences not $\time_2$ (as in \reffig{fig:zpr_ring_no_failures}), but $\time'_2$.
From the point of view of the original (not prime) fork, $\time_2$ is missing a sequence attestation.
From the point of view of the prime form, $\time'_3$ and $\time'_4$ are missing sequence attestations.
To know which interpretation is correct and which \seqsvclong s may have failed and need replacement, we need to decide whether $\time_2$, or $\time'_2$ is on the main chain.

\begin{figure}[h]
\centering
\vspace{-10pt}
\input{images/zpr_ring_with_fork}
\vspace{-5pt}
\caption{\sol\ with replacement ring sequence attestations and a fork.}
\label{fig:zpr_ring_with_fork}
\end{figure}

In \refalg{alg:omc} the \omc\ function definition relies on an order relation $\rightarrow$ between triads.\footnotemark
\footnotetext{In the context of \omc\ in \refalg{alg:omc}, we use the order relation on line \ref{alg:omc:findMinTriad} to find the successor triad with the lowest sequence attestation counter value.}
As we will see, it may be possible for a block and a timestamp attestation to exist on the main chain without forming a triad with a sequence attestation.
Accordingly, we define the order relation on timestamp attestations.
The timestamp attestation order relation relies on a \votes\ function, which returns the number of votes cast on a particular timestamp attestations.
Given two timestamp attestations $\time$ and $\time'$ we define a \emph{partial} order relation as
$$\height(\time) < \height(\time') \lor 
\left( \height(\time) = \height(\time') \land \votes(\time) > |\ring|/2 \right)
\implies
\time \rightarrow \time'.
$$
In turn, we define the $\votes(\time, \votectr, \depth, \seqs^*, \zpr, \blocks, \times, \seqs)$ function in \refalg{alg:votes}, where 
$\time$ is a timestamp attestation whose number of votes we want to count,
$\votectr$ is the current vote count,
$\depth$ is the remaining recursion depth,
$\seqs^*$ is a $\texttt{map}[\key^+]\seq$ of the most recently acknowledged sequence attestations for each \seqsvclong\ in $\ring$ at $\time$,
$\zpr$ is current main chain,
and $\blocks$, $\times$, and $\seqs$ is the set of known blocks, timestamp attestations, and sequence attestations.
Given the \votes\ function, we update the \omc\ function definition in \refalg{alg:omc_ring}.

\begin{minipage}{0.48\textwidth}
\begin{algorithm}[H]
\caption{$\votes(\time, \votectr, \depth, \seqs^*, \zpr, \blocks, \times, \seqs)$}
\label{alg:votes}
\begin{algorithmic}[1]
    \LComment{Count seq. attests. on $\time$ as votes}
    \For{$\seq^* \in \seqs^*$}
        \If{$\exists \seq \in \seqs ~|~ 
        \seq.\time^\uuid = \time.\uuid \land \seq.\time^\hash = \hash(\time) \land \\
        \mbox{~~}\seq.\ctr = \seq^*\!\!\!.\ctr+1 \land \check(\seq^*\!\!\!.\key, \seq) $}
            \State $\votectr \gets \votectr + 1$
            \State $\seqs^* \gets \seqs^* - \seq^*$
            \Comment{Count each sequencer once}
        \EndIf
    \EndFor

    \LComment{Return if found all the votes, or recursed enough}
    \If{$|\seqs^*| = 0 \lor \depth = 0 $}
        \State \Return $\votectr$
    \EndIf

    \LComment{Recurse into successors of $\time$}
    \For{$\block \in \blocks ~|~ \block.\time^\uuid = \time_\idxj.\uuid \land \block.\time^\hash = \hash(\time_\idxj)$}
        \LComment{Move up $\seqs^*$ by seq. attests. acknowledged in $\zpr$}
        \State $\seqs^\block \gets \seqs^*$
        \State $\block.\seqs \gets \textit{sort}(\block.\seqs, \seq.\ctr)$
        \Comment{Sort $\block.\seqs$ by counter value} 
        \For{$\seq \in \block.\seqs, \time \in \zpr ~|~ 
        \seq.\time^\uuid = \time.\uuid \land \seq.\time^\hash = \hash(\time) \land \\
        \mbox{~~}\seqs^\block[\seq.\key].\ctr + 1 = \seq.\ctr \land \check(\seqs^\block[\seq.\key].\key, \seq)$}
            \State $\seqs^\block[\seq.\key] \gets \seq$
        \EndFor
        
        \LComment{Recurse into successors of $\block$}
        \For{$\time \in \times ~|~ \time.\bytes = \langle \hash(\block), \block.\uuid \rangle$}
            \State $\votectr \gets \votectr + \votes(\time, \votectr, \depth - 1, \seqs^\block, \zpr, \blocks, \times, \seqs)$
        \EndFor
    \EndFor

    \State \Return $\votectr$

\end{algorithmic}
\end{algorithm}
\end{minipage}
\hfill
\begin{minipage}{0.48\textwidth}
\begin{algorithm}[H]
\caption{$\omc(\block_0, \ring, \key^+_\timesvc, \encl, \blocks, \times, \seqs)$}
\label{alg:omc_ring}
\begin{algorithmic}[1]

\State $\zpr \gets [~]$
\Comment{Start with $\block_0$ on main chain}

\LComment{Create a map of last used sequence attestations}
\State $\seqs^* \gets \varnothing$
\For{$\encl \in \ring$}
    \State $\seq \gets \varnothing$
    \State $\seq.\key \gets \decrypt{\key_\enclsvc^+}{\encl}.\key$
    \State $\seq.\ctr = -1$
    \State $\seqs^*[\seq.\key] \gets \seq$ 
\EndFor

\LComment{Add successor timestamp attestation if enough votes}
\State $\succs_\blocks \gets \left\{\block_0\right\}$
\Loop 
    \LComment{Find successor timestamp attestations}
    \State $\succs_\times \gets \left\{ \time \in \times ~|~ 
        \block \in \succs_\blocks \land \time.\bytes = \langle \hash(\block), \block.\uuid \rangle \land \validate(\key^+_\timesvc, \time) \right\}$
    \State $\succs_\blocks \gets \varnothing$

    \LComment{Does any $\time$ have enough votes}
    \For{$\time \in \succs_\times$}
        \If{$\votes(\time, 0, 100, \seqs^*, \zpr, \blocks, \times, \seqs) > |\ring|/2$}
            \LComment{Enough votes on only this $\time$}
            \State $\block \gets \block \in \blocks ~|~ \langle \hash(\block), \block.\uuid \rangle = \time.\bytes$
            \State $\zpr \gets \zpr \mdoubleplus \block \mdoubleplus \time$
            
            \LComment{Find successor blocks to $\time$}
            \State $\succs_\blocks \gets \left\{ \block \in \blocks ~|~ \block.\time^\uuid = \time.\uuid \land \block.\time^\hash = \hash(\time) \right\}$
            \LComment{Update $\seqs^*$ with acknowledged $\seq$ in $\block$}
            \For{$\seq \in \block.\seqs$}
                \If{$\seq.\ctr > \seqs^*[\seq.\key].\ctr$}
                    \State $\seqs^*[\seq.\key] \gets \seq$
                \EndIf
            \EndFor
        \EndIf
    \EndFor

    \LComment{Return if no $\time$ got enough votes to update $\succs_\blocks$}
    \If{$\succs_\blocks = \varnothing$}
        \State \Return $\zpr$
    \EndIf
\EndLoop

\end{algorithmic}












\end{algorithm}
\end{minipage}

%% file: images/write_ring.tex
\begin{tikzpicture}[node distance=2cm]
\node (user) [process] {\userlong};
\node (batchit) [process, below of=user] {\batchit};
\node (zipit) [process, below of=batchit] {\zipit};
\node (timestampit) [process, right=2.1cm of zipit] {\makecell[c]{Timestamp\\Service}};
\node (storeit) [process, below of=zipit] {\storeit};
\node (replicateit) [process, left=1.9cm of storeit] {\makecell[c]{Replication\\Service}};
\node (sequenceit) [process, right=2.1cm of storeit] {\sequenceit};
\node (ring) [process, right=2.1cm of sequenceit] {\makecell[c]{$\ring$}};

\draw [farrow] (user) -- node[anchor=west] {\ref{step:write:user->batchit}~$\tx$} (batchit);


\draw let \p1 = (batchit), \p2 = (replicateit) in
(\x2, \y1) node (br) [anchor=center] {};
\draw [farrow] (batchit) 
-- node[anchor=south] {\ref{step:write:batchit->replication}~$\merkle$} (br.center)
-- (replicateit)
;

\draw [farrow] (batchit) 
-- node[anchor=west] {\ref{step:write:batchit->zipit}~$\langle \hash(\merkle), \merkle.\uuid \rangle$} 
(zipit)
;


\draw [darrow] (zipit) -- 
node[anchor=south] {\ref{step:write:zipit->attest}~$\langle \hash(\block), \block.\time^\uuid \rangle$} 
node[anchor=north] {$\time$} 
(timestampit);

\draw [farrow] (zipit) -- 
node[anchor=west] {\ref{step:write:zipit->storeit}~$\langle \block, \time \rangle$} 
(storeit);

\draw [farrow] 
([yshift=0.0cm]storeit.west) 
-- 
node[anchor=south] {\ref{step:write:storeit->replication}~$\langle \block, \time \rangle$} 
([yshift=0.0cm]replicateit.east);

\draw [farrow] 
([yshift=0.0cm]storeit.east) 
-- 
node[anchor=south] {\ref{step:write_ring:storeit->seq}~$\langle \hash(\time), \time.\uuid \rangle$} 
([yshift=0.0cm]sequenceit.west);

\draw [darrow] 
(sequenceit.east) 
-- 
node[anchor=south] {\ref{step:write_ring:seq->ring}~$\langle \hash(\time), \time.\uuid \rangle$} 
node[anchor=north] {\ref{step:write_ring:ring->seq}~$\seq$} 
(ring.west);

\node (ctrl1) [below=0.5cm of storeit, anchor=center] {};
\node (ctrl2) [below right=0.5cm and 8cm of storeit, anchor=center] {}; 
\node (ctrl3) [right=8cm of batchit, anchor=center] {}; 
\draw[farrow] (storeit) 
-- (ctrl1.center) 
-- node [anchor=north] {\ref{step:write_ring:storeit->seq}~$\merkle.\uuid$} (ctrl2.center)
-- (ctrl3.center)
-- (batchit)
;

\draw [farrow] 
(sequenceit.south) 
to [out=240, in=300]
node[anchor=south] {\ref{step:write_ring:seq->repl}~$\seq$} 
([xshift=0.25cm]replicateit.south);

\draw [farrow] 
([xshift=-0.3cm]sequenceit.north) 
to [out=130, in=350]
node[anchor=north] {\ref{step:write_ring:repl->zipit}~$\seq$~~~~~} 
([yshift=-0.25cm]zipit.east);

\end{tikzpicture}

%% file: images/zpr_ring_no_failures.tex
\begin{tikzpicture}[node distance=1.5cm]
\node (b0) [block] {$\block_{0}$};
\node (t0) [block, above right = 1.5cm and 0.7cm of b0] {$\time_{0}$};
\node (b1) [block, below right= 1.5cm and 0.7cm of t0] {$\block_{1}$};
\node (t1) [block, above right= 1.5cm and 0.7cm of b1] {$\time_{1}$};
\node (b2) [block, below right= 1.5cm and 0.7cm of t1] {$\block_{2}$};
\node (t2) [block, above right= 1.5cm and 0.7cm of b2] {$\time_{2}$};
\node (b3) [block, below right= 1.5cm and 0.7cm of t2] {$\block_{3}$};
\node (t3) [block, above right= 1.5cm and 0.7cm of b3] {$\time_{3}$};
\node (b4) [block, below right= 1.5cm and 0.7cm of t3] {$\block_{4}$};
\node (t4) [block, above right= 1.5cm and 0.7cm of b4] {$\time_{4}$};

\node (s00) [seq_block, right= 0.7cm of b0] {$\seq^0_0$};
\node (s10) [seq_block, right= 0.7cm of b1] {$\seq^1_0$};
\node (s20) [seq_block, right= 0.3cm of b2] {$\seq^2_0$};
\node (s01) [seq_block, right= 0.3cm of b3] {$\seq^0_1$};
\node (s02) [seq_block, below right = 0.3cm and 1.2cm of b3] {$\seq^0_2$};
\node (s11) [seq_block, right= 0.7cm of b4] {$\seq^1_1$};

\draw [farrow]      (t0)  -- (b0);
\draw [seq_arrow]   (s00) -- (t0);
\draw [farrow]      (b1)  --  (s00);

\draw [farrow]      (b1)  -- (t0);
\draw [farrow]      (t1)  -- (b1);
\draw [seq_arrow]   (s10) -- (t1);
\draw [farrow]      (b3)  to [out=220, in=320] node[fill=backgroundcolor]  {$\seq^1_0.\sig$} (s10);

\draw [farrow]      (b2)  -- (t1);
\draw [farrow]      (t2)  -- (b2);
\draw [seq_arrow]   (s20) -- (t2);
\draw [farrow]      (b3)  -- node[fill=backgroundcolor]  {$\seq^2_0.\sig$} (s20);

\draw [farrow]      (b3)  -- (t2);
\draw [farrow]      (t3)  -- (b3);
\draw [seq_arrow]   (s01) -- (t3);
\draw [seq_arrow]   (s02) -- (t3);
\draw [farrow]      (b4)  -- (s01);
\draw [farrow]      (b4)  -- (s02);

\draw [farrow]      (b4)  -- (t3);
\draw [farrow]      (t4)  -- (b4);
\draw [seq_arrow]   (s11) -- (t4);

\end{tikzpicture}

%% file: images/zpr_ring_with_fork.tex
\begin{tikzpicture}[node distance=1.5cm]
\node (b0) [block] {$\block_{0}$};
\node (t0) [block, above right = 1.5cm and 0.7cm of b0] {$\time_{0}$};
\node (b1) [block, below right= 1.5cm and 0.7cm of t0] {$\block_{1}$};
\node (t1) [block, above right= 1.5cm and 0.7cm of b1] {$\time_{1}$};
\node (b2) [block, below right= 1.5cm and 0.7cm of t1] {$\block_{2}$};
\node (t2) [block, above right= 1.5cm and 0.7cm of b2] {$\time_{2}$};
\node (b3) [block, below right= 1.5cm and 0.7cm of t2] {$\block_{3}$};
\node (t3) [block, above right= 1.5cm and 0.7cm of b3] {$\time_{3}$};
\node (b4) [block, below right= 1.5cm and 0.7cm of t3] {$\block_{4}$};
\node (t4) [block, above right= 1.5cm and 0.7cm of b4] {$\time_{4}$};

\node (s00) [seq_block, right= 0.7cm of b0] {$\seq^0_0$};
\node (s10) [seq_block, right= 0.7cm of b1] {$\seq^1_0$};
\node (s01) [seq_block, right= 0.3cm of b3] {$\seq^0_1$};
\node (s02) [seq_block, below right = 0.3cm and 1.2cm of b3] {$\seq^0_2$};
\node (s11) [seq_block, right= 0.7cm of b4] {$\seq^1_1$};

\draw [farrow]      (t0)  -- (b0);
\draw [seq_arrow]   (s00) -- (t0);
\draw [farrow]      (b1)  --  (s00);

\draw [farrow]      (b1)  -- (t0);
\draw [farrow]      (t1)  -- (b1);
\draw [seq_arrow]   (s10) -- (t1);
\draw [farrow]      (b3)  to [out=220, in=320] node[fill=backgroundcolor]  {$\seq^1_0.\sig$} (s10);

\draw [farrow]      (b2)  -- (t1);
\draw [farrow]      (t2)  -- (b2);

\draw [farrow]      (b3)  -- (t2);
\draw [farrow]      (t3)  -- (b3);
\draw [seq_arrow]   (s01) -- (t3);
\draw [seq_arrow]   (s02) -- (t3);
\draw [farrow]      (b4)  -- (s01);
\draw [farrow]      (b4)  -- (s02);

\draw [farrow]      (b4)  -- (t3);
\draw [farrow]      (t4)  -- (b4);
\draw [seq_arrow]   (s11) -- (t4);

\node (b2') [block, below= 3.3cm of b2] {$\block'_{2}$};
\node (t2') [block, above right= 1.5cm and 0.7cm of b2'] {$\time'_{2}$};
\node (b3') [block, below right= 1.5cm and 0.7cm of t2'] {$\block'_{3}$};
\node (t3') [block, above right= 1.5cm and 0.7cm of b3'] {$\time'_{3}$};
\node (b4') [block, below right= 1.5cm and 0.7cm of t3'] {$\block'_{4}$};
\node (t4') [block, above right= 1.5cm and 0.7cm of b4'] {$\time'_{4}$};

\node (s20) [seq_block, right= 0.7cm of b2'] {$\seq^2_0$};

\draw [farrow]      (b2')  to [out=100, in=290]  (t1);

\draw [farrow]      (t2')  -- (b2');

\draw [farrow]      (b3')  -- (t2');
\draw [farrow]      (t3')  -- (b3');
\draw [seq_arrow]   (s20) --  (t2');
\draw [farrow]      (b3')  -- (s20);

\draw [farrow]      (b4')  -- (t3');
\draw [farrow]      (t4')  -- (b4');

\end{tikzpicture}